\documentclass{jfm}
\usepackage{graphicx}
\usepackage{epstopdf, epsfig}
\usepackage{blkarray}
\usepackage{stmaryrd}
\SetSymbolFont{stmry}{bold}{U}{stmry}{b}{n}

\usepackage{natbib}
\usepackage{amsmath,amssymb,bm,bbm}
\usepackage[utf8]{inputenc}
\usepackage[T1]{fontenc} % Required for accented characters
\usepackage{commath}
\usepackage{hyperref}
\usepackage{braket}
\newcommand\Ray{\mbox{\textit{Ra}}} % Rayleigh number
\newcommand\Nu{\mbox{\textit{Nu}}} % Nusselt number
\newcommand\Pe{\mbox{\textit{Pe}}} % Peclet number
\newcommand\slsD{\mathsfbi{D}} % for D
\newcommand{\Id}{\mathsfbi{I}} % Identity matrix
\newcommand{\Ze}{\mathsfbi{0}} % Zero matrix
\newcommand{\ii}{\mathrm{i}}
\newcommand{\ee}{\mathrm{e}}

\newcommand{\mbf}[1]{\mathbf{#1}}

\newcommand{\sbf}[1]{{\bm #1}}

\DeclareMathOperator{\sgn}{sgn}

\shorttitle{Convection with phase change at boundaries}
\shortauthor{S. Labrosse,  A. Morison, R. Deguen and T. Alboussi\`ere}

\title{Rayleigh-Bénard convection in a creeping solid with melting and
  freezing at either or both its horizontal boundaries}

\author{Stéphane Labrosse
  \corresp{\email{stephane.labrosse@ens-lyon.fr}},
  Adrien Morison,
  Renaud Deguen \and Thierry Alboussière}

\affiliation{ %\aff{1}
  Université de Lyon, ENSL, UCBL, CNRS, LGL-TPE, 46
  allée d’Italie, F-69364 Lyon, France }

\begin{document}
\maketitle
\begin{abstract}
  Solid state convection can take place in the rocky or icy mantles of
  planetary objects and these mantles can be surrounded above or below
  or both by molten layers of similar composition. A flow toward the
  interface can proceed through it by changing phase. This behaviour
  is modeled by a boundary condition  taking into account the competition between viscous
    stress in the solid, that builds topography of the interface with
    a timescale $\tau_\eta$, and convective transfer of the latent
    heat in the liquid from places of the boundary where freezing
    occurs to places of melting, which acts to erase topography, with
    a timescale $\tau_\phi$. The ratio $\Phi=\tau_\phi/\tau_\eta$  controls
  whether the boundary condition is the classical non-penetrative one
  ($\Phi\rightarrow \infty$) or allows for a finite flow through the
  boundary (small $\Phi$). We study Rayleigh-Bénard convection in a
  plane layer subject to this boundary condition at either or both its
  boundaries using linear and weakly non-linear analyses. When both
  boundaries are phase change interfaces with equal values of $\Phi$,
  a non-deforming translation mode is possible with a critical
  Rayleigh number equal to $24\Phi$. At small values of $\Phi$, this
  mode competes with a weakly deforming mode having a slightly lower
  critical Rayleigh number and a very long wavelength,
  $\lambda_c\sim 8\sqrt{2}\pi/ 3\sqrt{\Phi}$. Both modes lead to very
  efficient heat transfer, as expressed by the relationship between
  the Nusselt and Rayleigh numbers. When only one boundary is subject
  to a phase change condition, the critical Rayleigh number is
  $\Ray_c=153$ and the critical wavelength is $\lambda_c=5$. The
  Nusselt number increases about twice faster with Rayleigh number
  than in the classical case with non-penetrative conditions and the
  average temperature diverges from $1/2$ when the Rayleigh number is
  increased, toward larger values when the bottom boundary is a phase
  change interface.
\end{abstract}

\begin{keywords}
  Mantle convection, Rayleigh-Bénard convection, Buoyancy driven instability, Solidification/melting
\end{keywords}

\section{Introduction}
\label{sec:introduction}
Rayleigh-Bénard convection is one of the main heat transfer mechanisms
in natural sciences, responsible for most of the dynamics of the
atmosphere and oceans \citep{Pedlosky}, plate tectonics
\citep{Schubert_etal01}, dynamo action in planetary cores
\citep{Roberts_King2013}. It is also one of the most generic example of
pattern formation mechanism in fluid dynamics \citep[e.g.][]{Cross_Hohenberg93,Manneville2004}
and has therefore attracted a lot of attention for a century since the
work of Lord Rayleigh \citep{Rayleigh16}. However, the mathematical
and experimental studies of Rayleigh-Bénard convection have usually
considered boundary conditions that are not fully relevant to the natural
systems that justified them, their horizontal surfaces being generally
considered as subjected to no-slip or free-slip boundary
conditions. The former is valid for convection experiments in a tank
and for the natural fluids bounded by much more viscous enveloppes,
like the liquid cores of terrestrial planets and the bottom of the
ocean. The latter is often considered as an approximation for a
free-surface condition, as applies to a fluid bounded by a much less
viscous one. This is in particular the case of the solid
  planetary mantles that, on long timescales, behave like very
  viscous fluids \citep[e.g.][]{Turcotte_Oxburgh67,McKenzie_etal74,Jarvis_McKenzie80}
  and are bounded below and above by liquid or gaseous layers. This approximation
neglects the effect of the topography on convection and some studies
have been devoted to the modeling of these effects, which can be
dramatic when it is associated to, for example, intense volcanism in
hot planets \citep{Monnereau_Dubuffet02,Ricard_etal2014}.

In the present paper, we consider the effects of having horizontal
boundaries at which a solid-liquid phase change occurs on Rayleigh-Bénard
convection in the creeping solid, that has an infinite Prandtl number
\citep{Schubert_etal01}. For simplicity, we consider a
  Newtonian fluid with a uniform high viscosity, neglecting the effects
  of more complex rheologies
  \citep[e.g.][]{Parmentier78,Christensen_Yuen1989,Davaille_Jaupart93,Tackley2000,Bercovici_Ricard2014},
that is bounded by a low viscosity liquid of the same composition as
the convecting solid. The boundary between the liquid and the solid
consists of a phase change whose position is controlled by a Clapeyron
diagram relating pressure and temperature for phase equilibrium. In
the context of planetary interiors, the pressure is largely dominated
by the hydrostatic contribution and the interface is on average a
horizontal surface. The stress field 
  and associated dynamic pressure
due to the dynamics of the solid leads to deformation of
the interface with a viscous timescale $\tau_\eta$. 
The topography creates
variations of the thermal gradient on the liquid side which drives
a convective heat transfer in the liquid
acting to erase the topography by transporting the latent heat released by
freezing in topography lows to topography highs where melting occurs.
Other sources of motions in the liquid can also contribute to this
lateral heat transfer which happens on a timescale $\tau_{\phi}$, the expression of which being derived in section \ref{sec:cons-equat}.
The ratio of the
two timescales, $\Phi=\tau_\phi / \tau_\eta$, controls the behaviour
of the boundary. For a large value of $\Phi$, the topography is set by
the balance between the viscous stress in the solid and the buoyancy
of the topography, the phase change acting on a too long timescale to
affect the classical behaviour of the free surface. The buoyancy of
the topography is responsible for making the vertical velocity drop
to zero at the interface, which leads to an
effectively non-penetrating boundary condition. On the other hand, 
for low values of $\Phi$, the topography is erased by freezing and
melting at a rate greater than the one at which it is generated. The
removal of the associated buoyancy leads to a non-null velocity across
the interface.

This situation has already been considered in
the case of the dynamics of the Earth inner core
\citep{Alboussiere_etal2010,Monnereau_etal2010,Deguen_etal2013,Mizzon_Monnereau2013},
which is the solid iron sphere at the center of the liquid iron core
of the Earth. 
 \citet{Deguen_etal2013} have derived a general formulation
  of the boundary condition for arbitrary values of $\Phi$ and shown 
that the application of this boundary condition to a sphere
considerably changes the dynamics by decreasing the critical Rayleigh
number for the onset of thermal convection and allowing a new mode of
convection, the translation mode, where no deformation occurs in the
sphere, melting happens at the boundary of the advancing hemisphere
and freezing occurs at the trailing boundary. 

A similar situation arises for the ice shell of some satellites of
giant planets in the solar system which are believed to host a liquid
ocean below their ice layer
\citep{Pappalardo98,Khurana_etal1998,Gaidos_Nimmo00,Tobie_etal2003,Soderlund_etal2014,Cadek_etal2016}. 
Some of the largest of such satellites can also have a layer of high
pressure ices below their ocean
\citep{Grasset_etal2000,Sohl_etal03,Baland_etal2014}. Another
situation that implies such a melt-solid interface arises on all
terrestrial planets in their early stage when their silicate layer is
completely or largely molten owing to the high energy of their
accretion \citep{Solomatov2007,Elkins-Tanton2012}. Convection can
start in the solid mantle during its crystallisation from the magma
ocean, while a liquid layer persists above and/or below
\citep{Labrosse_etal07}. It is therefore interesting to consider convection in a layer,
not a full sphere, when a phase change boundary condition applies at
either or both its horizontal boundaries.

\citet{Deguen2013} performed such a study in the case of a spherical shell with a central
gravity linearly varying with radial position and showed that, again,
a translation mode is possible and favoured in the linear stability
analysis if both the upper and lower boundaries allow an easy phase
change, that is if each has a low value of the $\Phi$ parameter. The
purpose of the present paper is to extend the analysis to the plane
layer situation and perform the linear stability and weakly non-linear
analysis as a function of the phase change parameters of both
horizontal boundaries.

The boundary conditions are  presented in  section~\ref{sec:cons-equat},
section~\ref{sec:translation-mode} presents the translation mode of
convection, section~\ref{sec:non-transl-modes} presents the linear and
weakly non-linear analysis in the case when both horizontal boundaries
have the same value of the phase change parameter and
section~\ref{sec:solutions-with-only} shows the case when phase change
is only allowed on one boundary.

\section{Conservation equations and boundary conditions}
\label{sec:cons-equat}
\begin{figure}
  \centering
  \includegraphics[width=0.8\textwidth]{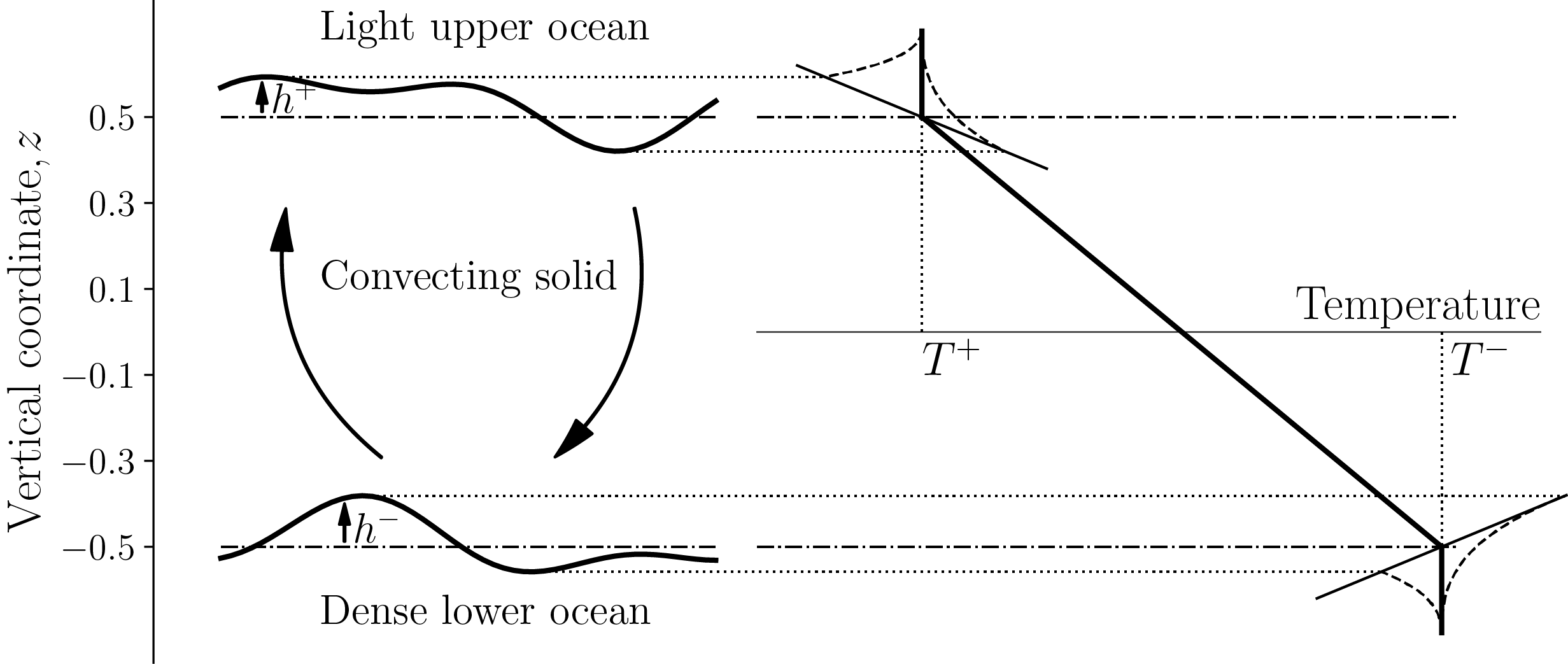}
  \caption{Definition of the topography (exaggerated here for clarity) and the temperature for the
    boundary conditions. The dash-dotted lines are the
    reference positions for the conductive motionless solutions of the
    top and bottom boundaries. The right panel shows the reference
    temperature profile (thick solid line) intersecting the melting temperature at
  the top and bottom (thin solid lines) at temperatures $T^+$ and $T^-$,
  respectively. Lateral variations of the topography make the
  intersection deviate laterally in temperature. Representative
  temperature profiles in the liquid sides are shown as dashed
  lines. In the context of planetary applications, the
    temperature profiles should be interpreted as deviations from the
    isentropic reference.
  }
  \label{fig:topo}
\end{figure}
We consider a layer of creeping solid that behaves like
  a Newtonian fluid on long timescales and that is bounded above
  or below or both by a liquid related to the solid by a phase change
  (fig.~\ref{fig:topo}). The temperature field at rest is solution of
  the thermal conduction problem with temperatures at the boundaries,
  $T^+$ at the top and $T^-$ at the bottom, that each equals the
  melting temperature $T_m$ at the relevant pressure. Pressure, in the
  context of planetary interiors, is largely dominated by the
  hydrostatic part. The melting temperature therefore mainly depends
  on the vertical coordinate. The possibility of crossing the melting
  temperature at both the top and bottom of our computational domain
  requires either a non-linear dependence of $T_m$ on pressure or,
  more easily, a compositional difference between the solid and both
  upper and lower liquid layers \citep{Labrosse_etal07}. For simplicity
  here, we do not consider the dynamical effects of compositional
  variations. The vertical dependence of the melting temperature is
  linearised around the reference positions of the boundaries, owing
  to the smallness of their topographies compared to the total
  thickness of the layer, $d$.

The conduction temperature profile that is used as reference writes
\begin{equation}
  \label{eq:T-cond}
  T_0 = \frac{T^+ + T^-}{2} + \frac{z}{d}\left(T^+-T^-\right),
\end{equation}
the reference for the vertical position $z$ being at the
center of the domain. Deviations from the conduction temperature
profiles are made dimensionless using $\Delta T = T^--T^+$ as reference and
denoted by $\theta$. In the following, superscripts $^+$ and $^-$ are
used for quantities pertaining to the top and bottom boundaries,
respectively, and omitted in equations that apply to both boundaries.

  The crossing positions of the conduction solution with the melting
  temperature at the top and bottom are used as reference around which
  a topography height, $h^+$ and $h^-$, is defined for each boundary,
  respectively (fig.~\ref{fig:topo}). These topographies can have either
  sign, positive upward, and need not average to 0, as will be shown below. At
  each phase change interface, two thermal boundary conditions are
  necessary to account for the moving boundary \citep{Crank}. The
  temperature must equal the phase change temperature and the heat
  flux discontinuity across the interface must balance the release or
  consumption of latent heat, $L$ (Stefan condition). 
 The two thermal boundary
conditions write
\begin{align}
  \label{eq:T-BC}
  T(h) &= T_m(h),\\
  \label{eq:q-BC}
  \rho_s L v_\phi &= \llbracket q \rrbracket,
\end{align}
with $v_\phi$ the freezing rate, $\rho_s$ the density
  of the solid and $\llbracket q \rrbracket$ the heat flux difference
  between the liquid and the solid sides. These boundary conditions
  apply to the deformed interface and need to be projected to the
  reference level that is used as boundary for the computation
  domain. Developing equation~\eqref{eq:T-BC} to first order in
  $h$ gives
\begin{equation}
  \label{eq:BC-T0}
  T\left(\pm\frac{d}{2}\right) = T^\pm + \left(\dpd{T_m^\pm}{z} - \dpd{T_0}{z}\right) h^\pm.
\end{equation}

  In dimensionless form,
  equation~\eqref{eq:BC-T0} writes
\begin{equation}
  \label{eq:BC-Theta0}
  \theta\left(\pm\frac{1}{2}\right) = \left(1 + \frac{d}{\Delta T}\dpd{T_m^\pm}{z}\right) \frac{h^\pm}{d}.
\end{equation}
In the following, we
  assume $h^\pm/d$ to be small
and we apply

\begin{equation}
  \label{eq:BC-T}
  \theta=0, \qquad z=\pm \frac{1}{2}.
\end{equation}

 Turning to the second thermal boundary condition, the
  discontinuity of heat flow on the right-hand-side of
  equation~\eqref{eq:q-BC} is assumed to be dominated by the
  convective heat flow on the low viscosity liquid side,
  $ f \sim \rho_lc_{pl}u_l\delta T_l$, with $\rho_l$ and $c_{pl}$ the
  density and heat capacity of the liquid, $u_l$ the characteristic
  liquid velocity and $\delta T_l$ the temperature difference between
  the boundary and the bulk of the liquid.  This difference results
  from variations of the topography (fig.~\ref{fig:topo}) and the
  vertical gradient of the melting temperature so that
\begin{equation}
  \label{eq:convective-flux-liq}
  f \sim - \rho_l c_{pl} u_l \left|\dpd{T_m}{z}\right| h.
\end{equation}
 The temperature difference
  $h \partial{T_m}/ \partial {z}$ is negligible on the solid side, but
  crucial for the convective heat flux on the liquid side.
  Fig.~\ref{fig:topo} shows as dashed lines the typical local
  temperature profiles on the liquid side of each boundary for
  topography highs and lows, indicating that the implied lateral
  variations of heat flux density should lead to melting of regions
  where the solid protrudes in the liquid and freezing in depressed
  regions, tending toward erasion of the topography. This behaviour is
  ensured by the anti-correlation of $f$ and $h$ in
  equation~\eqref{eq:convective-flux-liq}, independently of the sign
  of $\pd{T_m}{z}$, and this applies to both top and bottom
  boundaries.  The case of $\pd{T_m}{z} < 0$ depicted here for the top
  boundary is the most usual and the opposite case depicted here for
  the bottom boundary is encountered for water. Note, however, that in
  the context of planetary applications, the temperature considered
  here in the liquid layers and depicted on fig.~\ref{fig:topo} is in
  fact the deviation from the reference isentropic temperature profile
  \citep{Jeffreys1930,Deguen_etal2013} and the pressure derivative of
  the actual melting temperature needs not be negative for having a
  liquid underlying the solid layer.
  Assuming that the convective heat flow on the liquid side dominates
  the right hand side of equation~\eqref{eq:q-BC}, we write 

\begin{equation}
  \label{eq:phase-change-rate-flux}
  \rho_s L v_\phi \sim - \rho_l c_{pl} u_l \left|\dpd{T_m}{z}\right| h.
\end{equation}

  The freezing rate is related to the vertical velocity $w$ across the
  boundary and the rate of change of the topography as

\begin{equation}
  \label{eq:phase-change-rate}
  v_\phi^\pm = \pm \dpd{h\pm}{t} \mp w.
\end{equation}

  Combining with equation~\eqref{eq:phase-change-rate-flux} gives

\begin{equation}
  \label{eq:uz}
  w \mp \dpd{h}{t} =  \frac{\rho_l c_{pl} u_l}{\rho_s L}\left|\dpd{T_m}{z}\right| h
  \equiv \frac{h}{\tau_\phi},
\end{equation}
with
\begin{equation}
  \label{eq:tau-phi}
  \tau_\phi= \frac{\rho_s L}{\rho_l c_{pl} u_l \left|\pd{T_m}{z}\right|}
\end{equation}
  the characteristic phase change timescale for changing the
  topography by transferring latent heat from regions where it is
  released to places where it is consumed.  $u_l$ depends on the
  dynamics of the liquid which is not solved in this paper. The
  uncertainty in this quantity as well as the scaling coefficients
  implied by the $\sim$ sign in
  equations~\eqref{eq:convective-flux-liq}
  and~\eqref{eq:phase-change-rate-flux} are all combined to make
  $\tau_\phi$ the control parameter in our study.

Across the
boundaries, the total traction  must be continuous. Assuming
that the topography is small (i.e. the horizontal gradient of $h^\pm$
is small compared to 1, $|\nabla_h h^\pm|\ll 1$), this writes
for the vertical component
\begin{equation}
  \label{eq:stress}
  -P_s(h^\pm)+2\eta\dpd{w}{z}=-P_l(h^\pm)
\end{equation}
where $P$ is total pressure, $s$ and $l$ are for the solid and liquid
sides, respectively, 
  and $\eta$ is the dynamic viscosity of the
  solid. The total pressure on the solid and
liquid sides is split into its
hydrostatic part, $P(0)-\rho_{s,l}gh^\pm$ ($z=0$ being the reference
for $h$ at each boundary) and the dynamic part $p$. On the liquid
side, viscous stress and pressure fluctuations are neglected. With
these assumptions, we get 
\begin{equation}
  \label{eq:stress-2}
  -p+(\rho_s-\rho_l^\pm)gh^\pm+2\eta\dpd{w}{z}=0.
\end{equation}
Note that the density difference across the phase change boundary,
$\Delta\rho^\pm=\rho_s-\rho_l^\pm$, takes different signs at the top
and bottom since the solid must be denser than the overlying liquid but less
dense that the underlying one. Therefore $\Delta\rho^+ > 0$ and
$\Delta\rho^- < 0$.

The topography at each boundary is produced as a result of total stress
in the solid, with a typical timescale
$\tau_\eta=\eta/|\Delta\rho^\pm| g d$ \citep[the post-glacial rebound
timescale, ][]{Turcotte_Schubert01}, and erased by
  melting and freezing, as discussed above, with a timescale
  $\tau_\phi$. Both timescales are generally much shorter than
  the timescale for convection in the whole domain, so that we assume
  that the topography adjusts instantaneously to the competition
  between viscous stress and phase change. Therefore, we neglect
  $\pd{h}{t}$ in equation~\eqref{eq:uz} and combining it with
  equation~\eqref{eq:stress-2} to eliminate $h^\pm$, we get
\begin{equation}
  \label{eq:combined-BC}
  -p+\Delta\rho^\pm g \tau_\phi^\pm w + 2\eta\dpd{w}{z}=0.
\end{equation}

Introducing the phase change dimensionless number \citep{Deguen_etal2013,Deguen2013}
\begin{equation}
  \label{eq:phase-number}
  \Phi^\pm=\frac{\tau_{\phi^\pm}|\Delta\rho^\pm| g d}{\eta}
\end{equation}
equation~\eqref{eq:combined-BC}
takes the dimensionless form
\begin{equation}
  \label{eq:BC-wp}
  \pm\Phi^\pm w +2\dpd{w}{z} -p = 0, \qquad z=\pm \frac{1}{2}.
\end{equation}
$\Phi^\pm$ is the ratio of the phase change timescale to the viscous
deformation time scale. For large values of this parameter, the
boundary condition~\eqref{eq:BC-wp} reduces to the usual
non-penetration condition, $w=0$, while for small values it allows a
non zero mass flow through the boundary. The physical interpretation
is straightforward: if $\tau_\eta\ll \tau_{\phi^\pm}$, topography
evolves without the possibility of the phase change to happen and is limited by its
own weight that has to be supported by viscous stress in the solid. In
practice, this means that the flow velocity goes to zero at the free
interface and is very small at the reference boundaries $z=\pm 1/2$, which is
usually modeled as a non-penetrating boundary. In
the other limiting case, $\tau_\eta\gg \tau_{\phi^\pm}$, topography is
removed by phase change as fast as it is created by viscous stresses
and this allows a flow across the boundary. 

The liquid is assumed inviscid and therefore exerts no shear stress on
the convecting solid. The topography of the boundary is assumed to be
small and
we approximate the
  horizontal component of the continuity condition for traction by  a free-slip boundary condition at both horizontal
boundaries,
\begin{equation}
  \label{eq:BC-freeslip}
  \dpd{u}{z}+\dpd{w}{x}=0, \qquad z=\pm \frac{1}{2}.
\end{equation}

The dimensionless equations for the conservation of momentum, mass and energy are
written in the classical Boussinesq approximation as 
\begin{align}
  \label{eq:momentum}
  \frac{1}{\Pran}\left(\dpd{\sbf{v}}{t}+\sbf{v}\sbf{\cdot}\sbf{\nabla}
    \sbf{v}\right) & = -\sbf{\nabla}p + \nabla^2\sbf{v} +\Ray \theta
  \hat{\sbf{z}},\\
  \label{eq:mass}
  \bnabla\bcdot\boldsymbol{v} &= 0,\\
  \label{eq:energy}
  \dpd{\theta}{t}+\sbf{v}\sbf{\cdot}\sbf{\nabla}\theta &= w+\nabla^2\theta,
\end{align}
where $\Pran=\nu / \kappa$ is the Prandtl number, with $\nu$ and
$\kappa$ the momentum and thermal diffusivities, $\sbf{v}=(u, v, w)$ is the fluid velocity, $p$
is the dynamic pressure, $\Ray=\alpha \Delta T g d^3 / \kappa \nu$ is
the Rayleigh number, with $\alpha$ the thermal expansion
coefficient,

and $\hat{\sbf{z}}$ is the upward vertical unit vector. 
These equations have been made dimensionless using the
thickness of the layer $d$ as length scale and
the thermal diffusion time
$d^2/\kappa$ as timescale.

Since we are concerned here with
convection in solid, albeit creeping, layers, we will generally
consider the Prandtl number to be infinite in most of the calculations
below.

\section{The translation mode}
\label{sec:translation-mode}
The boundary condition~\eqref{eq:BC-wp} discussed in the previous
section permits a non-zero vertical velocity across the boundaries. If
both boundaries are semi-permeable (finite values of both $\Phi^+$ and
$\Phi^-$), the possibility of a uniform
vertical translation arises. This situation has been explored
systematically in the context of the dynamics of Earth's inner
core \citep{Alboussiere_etal2010,Deguen_etal2013,Mizzon_Monnereau2013}
and in spherical shells \citep{Deguen2013} but, in the case of a
spherical geometry, the horizontally average vertical velocity is
still null for a translation mode. Here we show that a translation
mode with a uniform vertical velocity also exists in the case of a plane
layer.

We search for a solution that is independent from the horizontal
direction and therefore only has a vertical velocity, $\sbf{v}=w
\hat{\sbf{z}}$. The mass conservation equation~\eqref{eq:mass} implies
that $w$ is independent of $z$ and we consider two situations, the
linear stability problem for which $w=W \mathrm{e}^{\sigma t}$ and the
steady state case for which $w$ is constant. Similarly, we can write
the temperature as $\theta(z, t)=\Theta(z) \mathrm{e}^{\sigma t}$ to study
the onset of convection in that mode, and $\theta$ as a function of
$z$ only at steady state and similar convention for pressure as $p$
and $P$.

\subsection{Linear stability analysis}
\label{sec:line-stab-analys}

The conservation equations~\eqref{eq:momentum}-~\eqref{eq:energy}
linearized around the hydrostatic state reduce to two equations
\begin{align}
  \label{eq:mom-trans-lin}
  \frac{\sigma}{\Pran}W&=-\Dif P + \Ray \Theta,\\
\label{eq:energy-trans-lin}
  \sigma \Theta &= W + \Dif^2\Theta,
\end{align}
with $\Dif\equiv\od{}{z}$. For neutral stability, $\sigma=0$, solving in
turn equation~\eqref{eq:energy-trans-lin} for $\Theta$ and
equation~\eqref{eq:mom-trans-lin} for $P$
subject to the boundary conditions~\eqref{eq:BC-T}
and~\eqref{eq:BC-wp} lead to
\begin{equation*}
  \left[\Ray-12\left(\Phi^++\Phi^-\right)\right]W=0.
\end{equation*}
A non-trivial solution for $W$ can then exist for
\begin{equation}
  \label{eq:Ra-crit-trans}
  \Ray = \Ray_c=12\left(\Phi^++\Phi^-\right),
\end{equation}
which is the condition for marginal stability of the translation
mode.

This system of equations can also be solved for a finite value of
$\sigma$ in order to relate it to
$\Ray$. Equation~\eqref{eq:energy-trans-lin} subject to boundary
conditions $\theta(\pm1/2)=0$ gives
\begin{equation}
\Theta = \frac{W}{\sigma} \left[  1 - 2 \frac{\sinh (\sigma^{1/2}/2)}{\sinh (\sigma^{1/2})} \cosh (\sigma^{1/2} z)   \right]
\end{equation}
Inserting this expression in Eq.~\eqref{eq:mom-trans-lin} and solving
for $P$, we obtain 
\begin{equation}
P = cst + \left(\frac{\Ray}{\sigma} - \frac{\sigma}{\Pran}  \right) W z - 2\Ray\, W \sigma^{-3/2}  
\frac{\sinh (\sigma^{1/2}/2)}{\sinh (\sigma^{1/2})} \sinh  (\sigma^{1/2} z).
\end{equation}
Using the boundary condition~\eqref{eq:BC-wp} at $z=1/2$ allows to
determine the integration constant, which gives 
\begin{equation}
\begin{split}
P = &\Phi^{+} W + \left(\frac{\Ray}{\sigma} - \frac{\sigma}{\Pran}  \right) W (z -1/2) \\ 
&- 2 \Ray\, W \sigma^{-3/2}  \frac{\sinh (\sigma^{1/2}/2)}{\sinh (\sigma^{1/2})} \left[ \sinh  (\sigma^{1/2} z) - \sinh  (\sigma^{1/2} /2) \right].
\end{split}
\end{equation}
Finally, using the boundary condition at $z=-1/2$, $-\phi^{-} W =
P(-1/2)$, gives, after rearranging, the following dispersion equation: 
\begin{equation}
0 = \frac{\sigma^{2} }{\Pran (\Phi^{+}+\Phi^{-})}+ \sigma + \frac{\Ray}{\Phi^{+}+\Phi^{-}} \left[ 2 \sigma^{-1/2} \frac{\cosh \sigma^{1/2} -1}{\sinh \sigma^{1/2}}  - 1 \right]. \label{Eq:Dispersion}
\end{equation}
An approximate solution for small $\sigma$ can be obtained by
developing the ratio of $\cosh$ and $\sinh$ functions to the second
order in $\sigma$, which gives
\begin{equation}
\sigma = \frac{10}{1+\frac{120}{\Pran\, \Ray}} \left( 1 - \frac{12(\Phi^{+}+\Phi^{-})}{\Ray}  \right).
\end{equation}
The critical Rayleigh number, obtained by setting $\sigma=0$, is the
same as that of Eq.~\eqref{eq:Ra-crit-trans}.
If $Gr_T \equiv \Pran\, \Ray$ (similar to the Grashof number but with
$\kappa$ in place of $\nu$) is large, the
expression for the growth rate reduces to
\begin{equation}
\sigma = 10 \left( 1 - \frac{12(\Phi^{+}+\Phi^{-})}{\Ray}  \right).
\end{equation}

\begin{figure}
\begin{center}
  \includegraphics[width=0.7\linewidth]{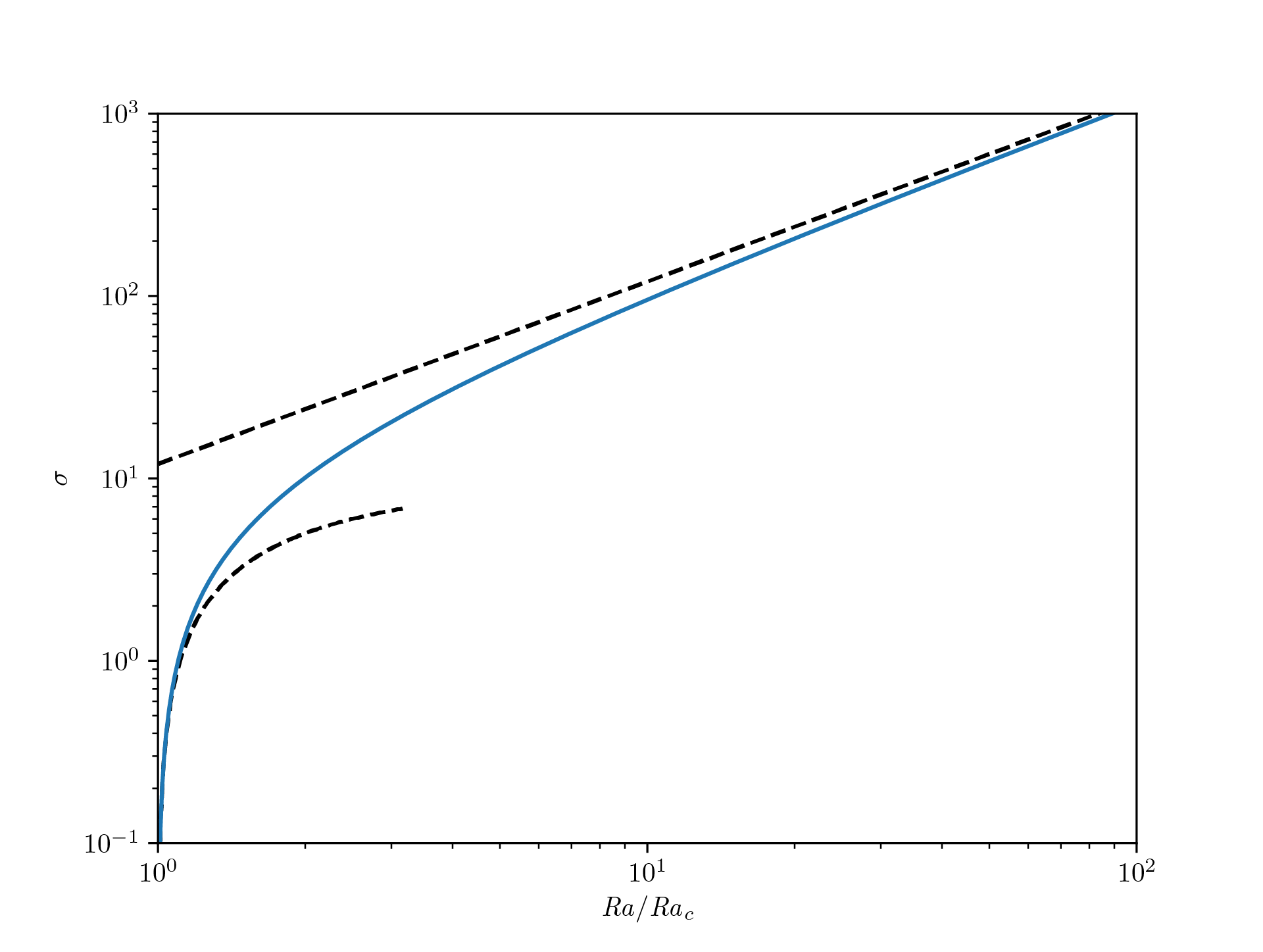}
\end{center}
\caption{Instability growth rate $\sigma$ as a function of $\Ray/\Ray_{c}$,
  for infinite $Gr_T$, as given by the numerical solution of the full
  dispersion relation (solid blue line), and by the small and large $\sigma$
  approximations (black dashed lines).}
\label{fig:growth-rate}
\end{figure}
 In the limit of a large $\sigma$, 
\begin{equation}
2 \sigma^{-1/2} \frac{\cosh \sigma^{1/2} -1}{\sinh \sigma^{1/2}} - 1 \rightarrow -1
\end{equation}
and the dispersion relation reduces to
\begin{equation}
0 = \frac{\sigma^{2} }{Gr_T}+   \frac{\Phi^{+}+\Phi^{-}}{\Ray} \sigma - 1.	\label{Eq:DispersionLargeSigma}
\end{equation}
The positive root is
\begin{equation}
\sigma = \frac{\Phi^{+}+\Phi^{-}}{\Ray}\frac{Gr_T}{2} \left[ \sqrt{1+\frac{4}{Gr_T}\left( \frac{\Ray}{\Phi^{+}+\Phi^{-}} \right)^{2}} - 1 \right]
\end{equation}
which reduces to
\begin{equation}
\sigma = \frac{\Ray}{\Phi^{+}+\Phi^{-}}
\end{equation}
in the limit of $\frac{1}{Gr_T}\left(
  \frac{\Ray}{\Phi^{+}+\Phi^{-}}\right)^{2}\ll 1$. The growth rate in
the large $Gr_T$ limit is plotted as function of $\Ray/\Ray_c$ on figure~\ref{fig:growth-rate}.

\subsection{Steady state translation}
\label{sec:steady-state-transl}

The steady state finite amplitude translation mode is solution of
\begin{align}
  \label{eq:mom-trans}
  0&=-\Dif p + \Ray \theta,\\
  \label{eq:energy-trans}
   w\Dif\theta&= w + \Dif^2\theta.
\end{align}
Solving first the energy balance equation~\eqref{eq:energy-trans}
subject to boundary conditions~\eqref{eq:BC-T} gives
\begin{equation}
  \label{eq:theta-trans}
  \theta=z+\frac{\cosh\left(\frac{w}{2}\right)-\mathrm{e}^{w z}}{2
    \sinh\left(\frac{w}{2}\right)}
\Rightarrow T= \frac{1}{2} +\frac{\cosh\left(\frac{w}{2}\right)-\mathrm{e}^{w z}}{2
    \sinh\left(\frac{w}{2}\right)}.
\end{equation}
Using the momentum balance equation~\eqref{eq:mom-trans} and the
boundary conditions~\eqref{eq:BC-wp} then gives
\begin{equation}
  \label{eq:W-Ra-trans}
  \left(\Phi^++\Phi^-\right)w=\Ray
  \left[
    \frac{\cosh\left(\frac{w}{2}\right)}{2\sinh\left(\frac{w}{2}\right)}-\frac{1}{w}
  \right].
\end{equation}
This transcendental equation relates the translation velocity $w$ to
the Rayleigh number. 

Close to onset, assuming the Péclet number, $|w|$, to be small,
equation~\eqref{eq:W-Ra-trans} can be developed  as function of
$(\Ray-\Ray_c)/\Ray_c$ to give to leading order
\begin{equation}
  \label{eq:W-trans-small}
  w=\pm 2\sqrt{15\frac{\Ray-\Ray_c}{\Ray_c}}.
\end{equation}
The corresponding temperature anomaly is
\begin{equation}
  \label{eq:theta-small-w}
  \theta = \frac{w}{8} \left(1 - 4 z^2\right) + O(w^2),
\end{equation}
showing that the temperature only differs from the conduction solution
by an amount proportional to the Péclet number.

\begin{figure}
  \centering
  \includegraphics[width=0.5\textwidth]{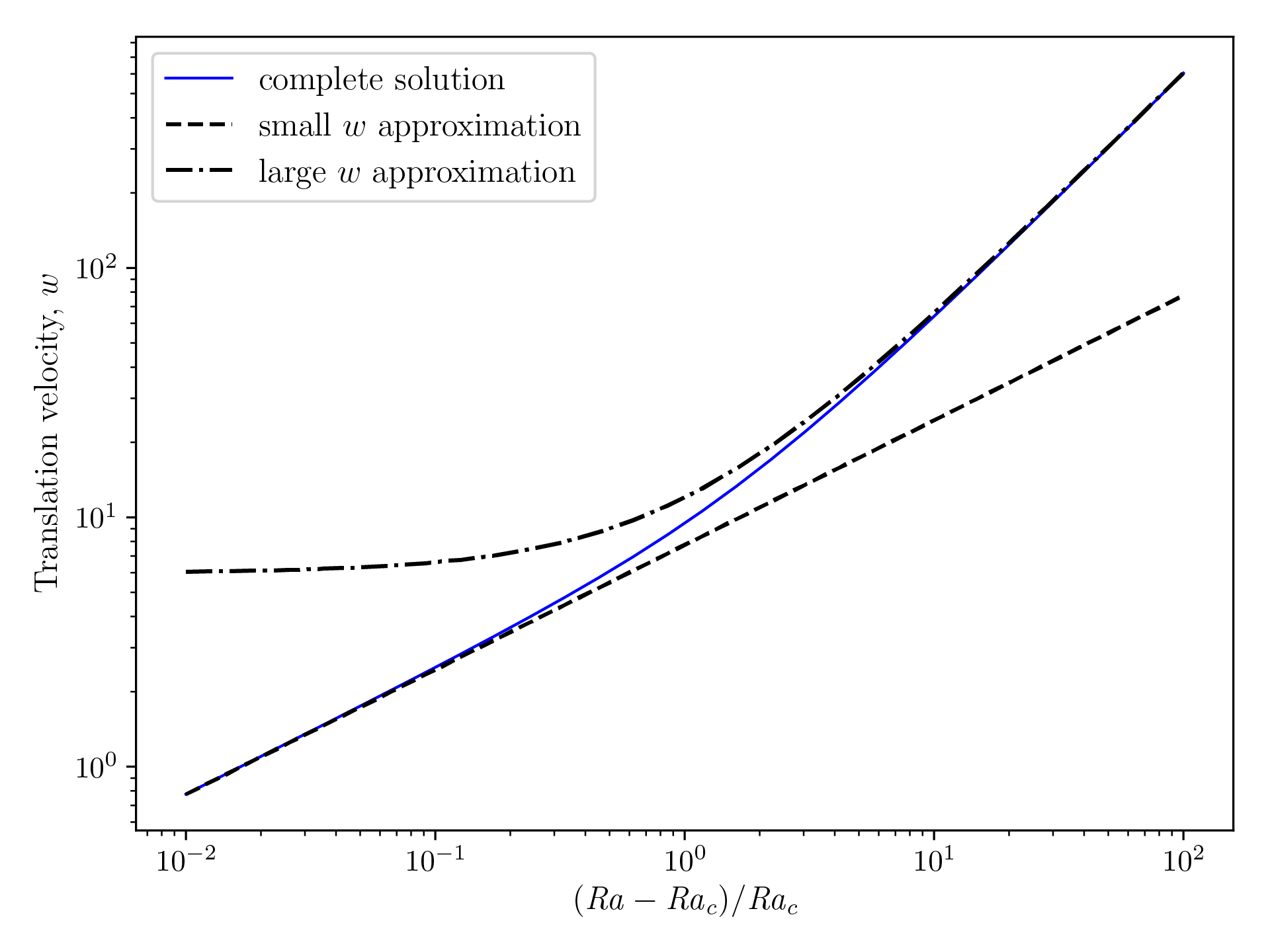}
  \caption{Finite amplitude velocity in the translation mode. The
    dashed line is the small velocity approximation given by
    eq.~(\protect\ref{eq:W-trans-small}), the dash-dotted line is the
    large velocity approximation given by eq.~(\ref{eq:W-Ra-large})
    and the solid line is the solution to the full equation~(\ref{eq:W-Ra-trans}).}
  \label{fig:w-translation}
\end{figure}
For a large Péclet number, $|w|\gg 1$, equation~\eqref{eq:W-Ra-trans} reduces to
\begin{equation}
  \label{eq:W-Ra-large}
  w\sim\pm\frac{\Ray}{2\left(\Phi^++\Phi^-\right)}=\pm\frac{6\Ray}{\Ray_c}. 
\end{equation}

Figure~\ref{fig:w-translation} shows how the translation velocity
$|w|$ depends on Rayleigh number, computed using the full
equation~(\ref{eq:W-Ra-trans}) and either the low or the large
velocity development. It shows that the transition
between the two regimes
happens for
$\Ray\sim 2\Ray_c$.

\begin{figure}
  \centering
  \includegraphics[width=0.5\textwidth]{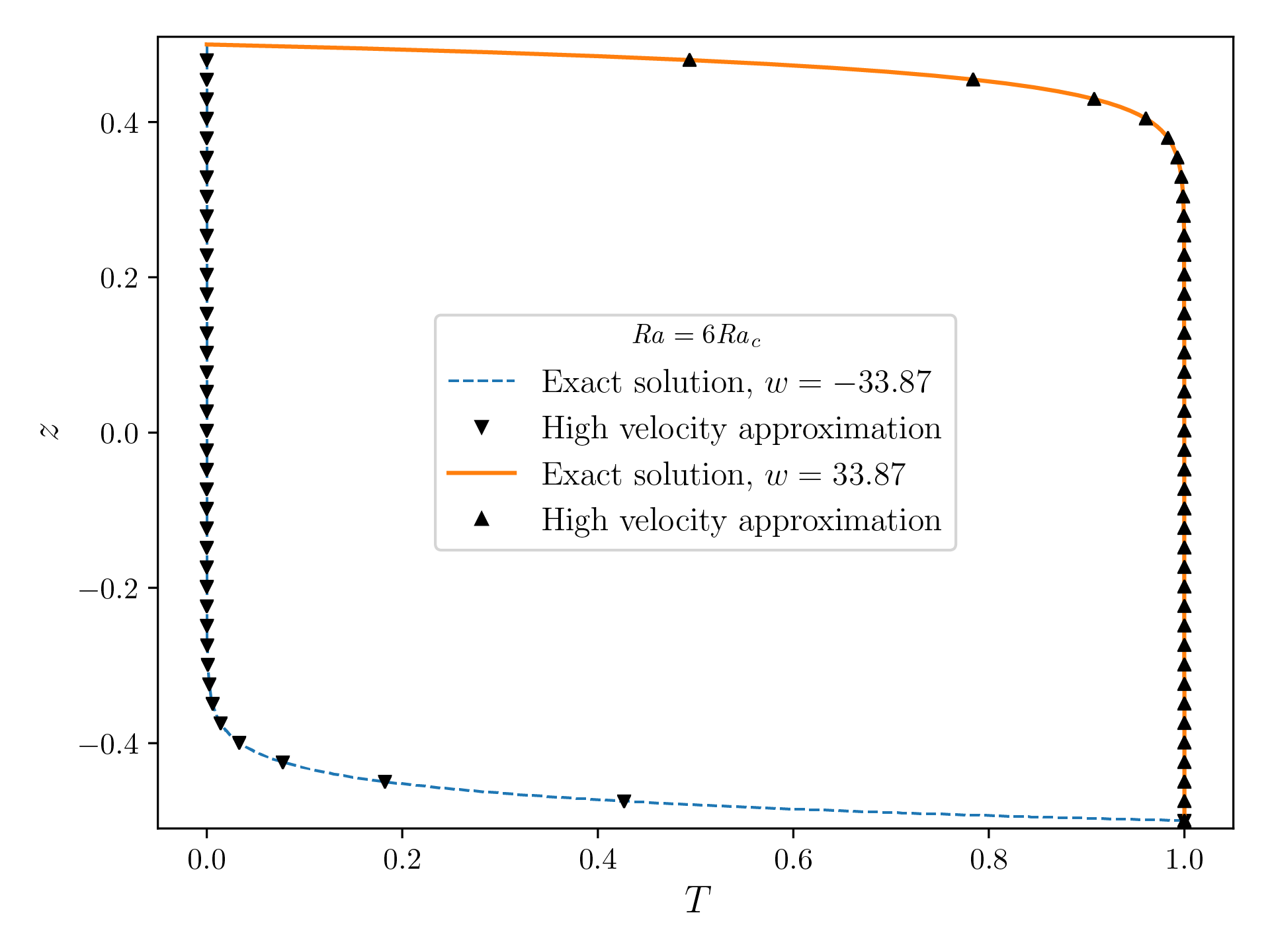}
  \caption{Temperature profile in the translation mode for
    $(\Ray-\Ray_c)/\Ray_c=5$. The solid (resp. dashed) line is for the
    ascending (resp. descending) mode calculated using the full
    equation~(\ref{eq:theta-trans}) and the up (resp. down) triangles
    are obtained using the approximate
    equation~(\ref{eq:theta-large-W}).}
  \label{fig:translation}
\end{figure}

In the high Péclet number regime, the temperature anomaly takes a simple form:
\begin{equation}
  \label{eq:theta-large-W}
  \theta\sim
  z+\sgn(w)\left[\frac{1}{2}-\mathrm{e}^{w(z-\sgn(w)/2)}\right] \Rightarrow 
T \sim \frac{1}{2}
\left[
  1+\sgn(w)
\right]-\sgn(w) \mathrm{e}^{w(z-\sgn(w)/2)}.
\end{equation}
The exponential in the last equation is negligible everywhere except
close to the upper boundary ($z=1/2$; resp. lower boundary, $z=-1/2$)
when $w\gg 1$ (resp. $w\ll -1$).  Therefore, the temperature is
essentially equal to that imposed at the boundary the fluid originates
from ($0$ at the top, $1$ at the bottom) and adjusts to that of the
opposite side in a boundary layer of thickness $\delta\sim 1/w$. In
dimensional units, $\delta$ is simply defined as the thickness that
makes the Péclet number around 1: $\Pe=w\delta/\kappa\sim
1$. Figure~\ref{fig:translation} shows the temperature  profiles for the
upward and downward translation modes computed both with the exact
(eq.~\ref{eq:theta-trans}) and approximate
(eq.~\ref{eq:theta-large-W}) expressions, showing that the
approximation is quite good. 

The steady state velocity given by equation~\eqref{eq:W-Ra-large} can
also be obtained from a simple physical argument. In the steady translation
regime, the (uniform) topography at each boundary is related to the
translation velocity and the phase change timescale by
\begin{equation}
  \label{eq:w-translation-tau}
  h^\pm=\tau_{\phi^\pm} w.
\end{equation}
In steady state, the excess (resp. deficit) weight of the cooler
(resp. warmer) solid
layer is balanced by the sum of pressure deviations from the
hydrostatic equilibrium at both boundaries as
\begin{equation}
  \label{eq:buoyancy-pressure}
  \alpha \rho_0 g\frac{\Delta T d}{2} = \Delta \rho^+ g h^+
  +\Delta \rho^- g h^-,
\end{equation}
where the temperature in the solid layer has been assumed uniform,
i.e. the contribution of the boundary layer to its buoyancy has been neglected.
This gives for the translation velocity
\begin{equation}
  \label{eq:w-translation}
  w=\frac{\alpha \rho_0 g\Delta T d}
  {2(\Delta \rho^+ g \tau_{\phi^+}+\Delta \rho^- g \tau_{\phi^-})}.
\end{equation}
In dimensionless form, this is exactly equation~\eqref{eq:W-Ra-large}.

It is also worth considering the heat transfer efficiency in the
translation mode. Equation~(\ref{eq:energy-trans}) can be integrated
to show that $w T - \Dif{T}$ is independent of $z$ and this implies
that $w = \Dif{T}(-\frac{1}{2})-\Dif{T}(\frac{1}{2})$, meaning that
the difference between the conductive heat fluxes across the horizontal
boundaries is equal to the advection by
translation. Figure~\ref{fig:translation} show that the heat flow
(Nusselt number $\Nu$) should be computed on the exit side, where a
boundary layer is produced:
\begin{equation}
  \label{eq:Nusselt1}
  \Nu= - \Dif{T} \left( \sgn{(w)}\frac{1}{2} \right) =
  |w|-\Dif{T}\left(-\sgn{(w)}\frac{1}{2}\right) = |w| + \frac{w\ee^{-|w|/2}}{2\sinh(w/2)}.
\end{equation}
The small and large $|w|$ limit cases give
\begin{align}
  \label{eq:Nu-small}
  \Nu &=1+\frac{|w|}{2} = 1+\sqrt{15\frac{\Ray-\Ray_c}{\Ray_c}},\\
  \label{eq:Nu-large}
  \Nu &= |w| = 6\frac{\Ray}{\Ray_c},
\end{align}
respectively. The large Rayleigh number behaviour is in striking
contrast to the situation encountered for standard Rayleigh-Bénard
convection for which $\Nu\sim\Ray^\beta$ with $\beta\sim 1/3$. 

\section{Non-translating modes with $\Phi^+=\Phi^-$}
\label{sec:non-transl-modes}
In this section, we consider the situation with values of the phase
change parameter of both boundaries equal, $\Phi\equiv \Phi^+=\Phi^-$.

\subsection{Linear stability}
\label{sec:linear-stability}

Non-translating solutions can be obtained using standard approaches for
the classical Rayleigh-Bénard problem. For the linear stability
problem, a solution using separation of variables is sought,
i.e. $u=U(z) \mathrm{e}^{\mathrm{i}k x} \mathrm{e}^{\sigma t}$ and similarly for
$w$, $p$ and $\theta$. Linearized equations~(\ref{eq:momentum})
to~(\ref{eq:energy}) reduce to
\begin{align}
  \label{eq:mass-linear}
  \mathrm{i} k U+D W &= 0,\\
  \label{eq:x-momentum-linear}
  \Pran\left[-\mathrm{i} k P + \left(\Dif^2-k^2\right) U\right] & = \sigma U,\\
  \label{eq:z-momentum-linear}
  \Pran\left[-\Dif P + \left(\Dif^2-k^2\right) W+\Ray \Theta\right] & = \sigma W, \\
  \label{eq:energy-linear}
  W+\left(\Dif^2-k^2\right) \Theta &= \sigma\Theta
\end{align}
since, at the linear stage, the problem is fully degenerate in terms
of orientation of the mode which can be taken as depending only on
$x$. These equations must be complemented by boundary conditions
applying at $z=\pm\frac{1}{2}$:
\begin{align}
  \label{eq:BC-u}
  \Dif U+\mathrm{i} k W & =0,\\
  \label{eq:BC-w}
  \pm\Phi^\pm W + 2\Dif W - P &=0, \\
  \label{eq:BC-theta}
  \Theta &= 0.
\end{align}
This forms a generalized eigenvalue problem that we solve using a
Chebyshev-collocation pseudo-spectral approach
\citep[e.g.][]{Canuto_etal1988,Guo_etal2012}. Given the
Chebyshev-Gauss-Lobatto nodal point $z_i=\cos\frac{i\upi}{N},
i=0...N$, in the interval $\left[-1, 1\right]$,
the values of the $z$-dependent mode functions at $z_i/2$ is
noted as $U_i$ for U and similarly for other variables. Division by
$2$ is required here to map the interval on which Chebyshev
polynomials are defined onto $\left[-\frac{1}{2},
  \frac{1}{2}\right]$. The $k^{th}$ derivative of each function  at
the nodal points is related to the nodal values of the function itself
by differentiation matrices:
\begin{equation}
  \label{eq:diff-matrix}
  \sbf{U}^{(k)}=\slsD^{(k)} \cdot \sbf{U}.
\end{equation}
The calculation of the differentiation matrices is done using a
\textit{python} adaptation\footnote{available at
  \href{https://github.com/labrosse/dmsuite}{https://github.com/labrosse/dmsuite}}
of DMSUITE \citep{Weideman_Reddy2000}. With these differentiation
matrices, the system of equations~\eqref{eq:mass-linear}
to~\eqref{eq:energy-linear} can be written as a generalized eigenvalue
problem of the form
\begin{equation}
  \label{eq:eigenproblem}
  \mathsfbi{L}\cdot \sbf{X} = \sigma\mathsfbi{R}\cdot\sbf{X}
\end{equation}
with $\sbf{X}=(\sbf{P}; \sbf{U}; \sbf{W}; \sbf{\Theta})^T$ the global
vertical mode 
  vector composed of the concatenation of vectors $P_i$, $U_i$, $W_i$
  and $\Theta_i$, 
 and $\mathsfbi{L}$ and $\mathsfbi{R}$ two
matrices representing the system with its boundary conditions. The
general structure of $\mathsfbi{L}$ reads as
\begin{equation}
  \label{eq:L-matrix}
  \setlength{\arraycolsep}{10pt}
  \renewcommand{\arraystretch}{1.3}
  \mathsfbi{L} = 
  % \left(
  \begin{blockarray}{ccccc}
    0:N & 0:N & 0:N & 1:N-1 & \\
    % mass
    \begin{block}{(cccc)l}
      \Ze & \ii k \Id & \slsD & \Ze  & \ 0:N\\
      % BC on U
      \Ze & \slsD & \ii k\Id & \Ze & \ 0\\
      % U
      -\Pran \ii k \Id & \Pran\left(\slsD^{(2)}-k^2\Id \right)
      & \Ze & \Ze & \ 1:N-1\\
      % BC on U
      \Ze & \slsD & \ii k\Id & \Ze  & \ N \\
      % BC on W
      -\Id & \Ze & \Phi^+\Id+2 \slsD & \Ze  & \ 0\\
      % W
      -\Pran \slsD & \Ze &
      \Pran\left(\slsD^{(2)}-k^2\Id\right)
      & \Pran \Ray \Id
      & \ 1:N-1\\
      % BC on W
      -\Id & \Ze & -\Phi^-\Id+2 \slsD & \Ze &\ N\\
      % T
      \Ze & \Ze & \Id &
      \left(\slsD^{(2)}-k^2\Id \right) &\ 1:N-1\\
    \end{block}
  \end{blockarray}
  % \right)
\end{equation}
with $\Id$ and $\Ze$ the identity and zero matrices, respectively.
The restrictions of line and column indices, indicated on the right and
top of the matrix respectively, are necessary to leave out the
boundary points from applications of equations~\eqref{eq:mass-linear} 
to~\eqref{eq:energy-linear} since these follow
equations~\eqref{eq:BC-u}-\eqref{eq:BC-theta} instead. For example, in
the second line of the matrix that represents
equation~\eqref{eq:BC-u}, only the first line (index $0$) of the
matrice $\Ze, \slsD, \ii k\Id$ and $\Ze$ are present.
Note that the
boundary values for the temperature are simply left out since the
Dirichlet boundary condition~\eqref{eq:BC-theta} is, in a collocation approach, naturally enforced
by removing the extreme Chebyshev points.

The $\mathsfbi{R}$ matrix contains ones on the diagonal
corresponding to the interior points of the equations for $\sbf{U}$,
$\sbf{W}$ and $\sbf{\Theta}$ and zeros elsewhere. When solving for an
infinite Prandtl number, which is the case below, the interior points
for the $\sbf{U}$ and
$\sbf{W}$ equations are also set to 0, leaving ones only for the interior points
of the $\sbf{\Theta}$ equation.
The resulting system
is singular and many eigenvalues are infinite, one for each zero in
the $\mathsfbi{R}$ matrix.
Filtering these spurious eigenvalues leaves us
with the relevant eigenvalues that are used to assess stability. For
any values of $\Phi^-$, $\Phi^+$ and $k$, the minimum value of $\Ray$
that makes the real part of one of the eigenvalues become positive is
the critical Rayleigh number for perturbations with that wavenumber. Minimizing $\Ray$ as function of $k$
gives the critical Rayleigh number for all infinitesimal perturbations. 
\begin{figure}
  \centering
  \includegraphics[width=\textwidth]{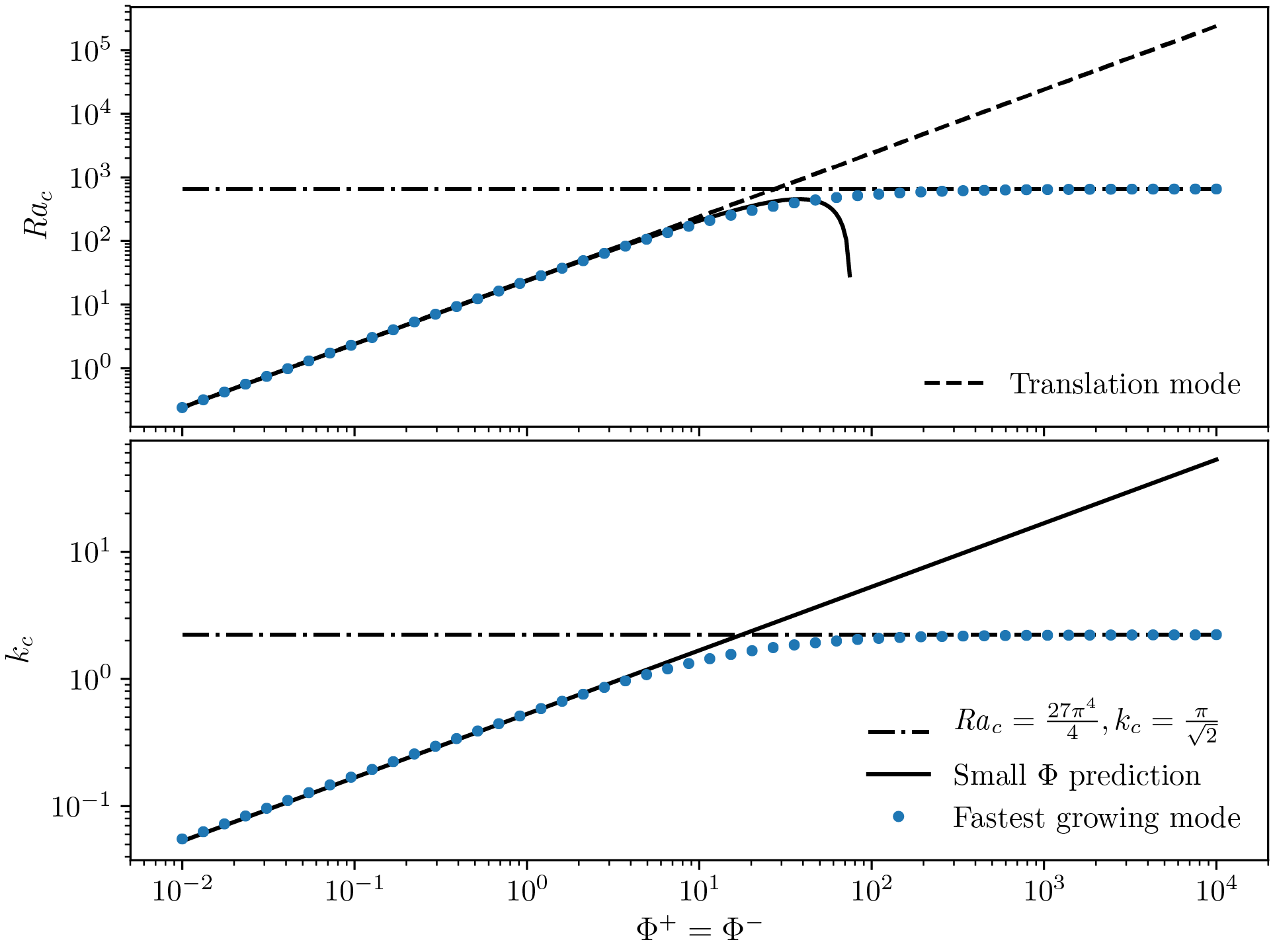}
  \caption{Critical Rayleigh number (top) and wavenumber (bottom) as function
    of the phase change numbers, both taken equal here. Filled circles
  are results of the calculation using the Chebyshev-collocation
  technique, the dash-dotted lines represent the classical
  $\Phi\rightarrow\infty$ limit, the dashed line in the upper panel
  represents the result for the translating mode
  (eq.~\ref{eq:Ra-crit-trans}) and the solid lines represent the
  small $\Phi$ leading order development.}
  \label{fig:Ra-crit-equal-phi}
\end{figure}
Figure~\ref{fig:Ra-crit-equal-phi} shows the evolution of the
critical Rayleigh number and the associated wavenumber as function of
the value of $\Phi^\pm$, both taken equal, $\Phi^+=\Phi^-=\Phi$. One can see that the
classical value derived by \citet{Rayleigh16} is recovered when
$\Phi\rightarrow\infty$, as expected. In the other limit,
$\Phi\rightarrow 0$, the critical Rayleigh number follows the
analytical expression obtained for the translation mode
(\S~\ref{sec:translation-mode}) while $k\rightarrow 0$, as
expected. 

The behaviour of the system in the limit of small $\Phi$ can be
obtained using a polynomial expansion of all the functions, both in
$z$ and $\Phi$. Specifically, considering the symmetry of the problem
around $z=0$, we write the temperature as
\begin{equation}
  \label{eq:temp-mode-poly}
  \Theta=\sum_{n=0}^Na_n z^{2n}.
\end{equation}
The Hermitian character of the linear problem (see
appendix~\ref{sec:self-adj}) ensures that $\sigma$ is real and,
therefore, $\sigma=0$ at onset. Then $W$ and $U$ can be
obtained using equations~\eqref{eq:energy-linear}
and~\eqref{eq:mass-linear}. Equations~\eqref{eq:x-momentum-linear}
and~\eqref{eq:z-momentum-linear} then provide two expressions for
$DP$ 
and their equality implies several equations, one for each polynomial
order considered. All the functions are developed to the same order
as the temperature, $2N$. Note that even if the definition of $\Theta$
for a given $N$ only requires $N+1$ coefficients $a_n$, the development of
the other profiles to the same order requires the inclusion of $a_n$
for values up to $n=N+2$ because of the derivatives in the linear system.
Using, for example, $N=2$ gives a pressure
gradient $DP$  that contains terms in
$z^{2n}, n=0..2$, and provides
therefore three independent equations for the equality between the two
expressions. With the symmetry considered here, the boundary
conditions~\eqref{eq:BC-u}--~\eqref{eq:BC-theta} bring three
additional equations for the coefficients $a_n$. 

Setting first $\Phi=0$ leads to a non trivial solution only for
$\Ray=0$ and $k=0$, the solution being equal to the low $\Phi$
development of the translation solution. To go beyond that,
each coefficient $a_n$ is itself developed as a polynomial of $\Phi$:
\begin{equation}
  \label{eq:a-phi}
  a_n=\sum_{j=0}^Ja_{n,j}\Phi^j.
\end{equation}
Similarly, the critical Rayleigh number $\Ray_c$ and the square of the
critical wavenumber $k^2$ are developed in
powers of $\Phi$:
\begin{equation}
  \label{eq:ra-phi}
  \Ray_c=\sum_{j=0}^Jr_j\Phi^j, \qquad k_c^2=\sum_{j=0}^JK_j\Phi^j.
\end{equation}
The three boundary conditions and the equations implied by the equality
of the two pressure expressions are then written and solved for increasing degrees in
the development in $\Phi$. In practice, we restrict ourselves to
$N=J=2$. At order 0 in $\Phi$, the set of linear equations can admit a
non-trivial solution only if the determinant of the implied matrix is
zero, which provides two possible values of $r_0$. The lowest one
admits a minimum, $r_0=0$, for $K_0=0$. This implies
$a_{2,0}=a_{3,0}=a_{4,0}=0$ and $a_{1,0}=-4 a_{0,0}$. At order 1 in
$\Phi$, we get directly that $a_{2,1}=a_{3,1}=a_{4,1}=0$, $a_{1,1}=-4
a_{0,1}$ and $r_1=24$ with no information on $K_1$. This is
however obtained at the next order where we find that $K_1=9/32$ minimizes
$r_2$, which is then $r_2=-81/256$. The order 2 coefficients are
also obtained as a function of $a_{0,0}$, which is the value of the
maximum of $\Theta$. These can then be used to determine the shape of
the different function $\Theta$, $W$, $U$ and $P$ for small values of
$\Phi$. To leading order in $\Phi$ we get
\begin{align}
  \label{eq:kc-phi}
  k_c & = \frac{3}{4\sqrt{2}}\sqrt{\Phi}\\
  \label{eq:Ray-phi}
  \Ray_c&=24 \Phi-\frac{81}{256}\Phi^2,\\
  \label{eq:Theta-lead}
  \Theta & = (1-4 z^2)\Theta_{max},\\
  \label{eq:W-lead}
  W &= 8 \Theta_{max},\\
  \label{eq:U-lead}
  U &= -3\mathrm{i}\sqrt{2\Phi} z \Theta_{max},\\
  \label{eq:P-lead}
  P &= \frac{z}{2}\left(39-64z^2\right)\Phi\Theta_{max}.
\end{align}
\begin{figure}
  \centering
  \includegraphics[width=\textwidth]{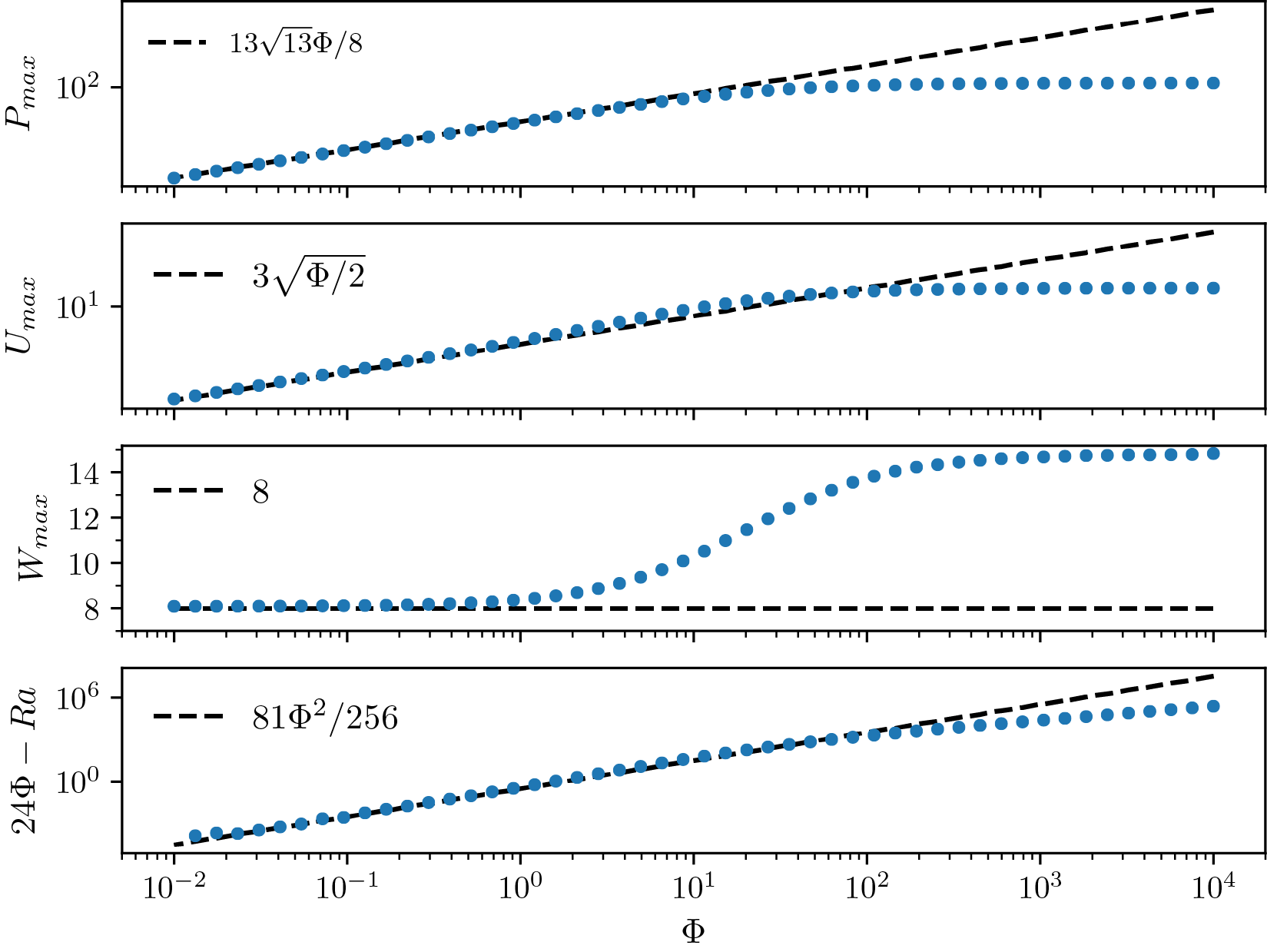}
  \caption{Variation of the maxima of profiles of $P$, $U$ and $W$ of
    the first unstable mode, that for $\Theta$ being set to 1, as a
    function of $\Phi$. The
    bottom panel shows the difference between $24\Phi$ and the
    critical Rayleigh number. On each plot, the solid circles are the
    results of the calculation using the Chebyshev-collocation method
    while the dashed lines are the low $\Phi$ predictions of
    equations~\protect\eqref{eq:Ray-phi} to~\protect\eqref{eq:P-lead}.}
  \label{fig:PhiModeMax}
\end{figure}
$\Theta_{max}=a_{0,0}$ is used to normalise all profiles. Note that
the shape of the temperature (eq.~\ref{eq:Theta-lead}) and vertical
velocity (eq.~\ref{eq:W-lead}) profiles are of order 0 in $\Phi$ and
are equal to their counterpart in the steady-state translation
solution (eq.~\ref{eq:theta-small-w}).
The small
$\Phi$ development of the solution to the linear problem can be
compared to the results obtained using the Chebyshev-collocation
method for cross-validation. The match between the mode profiles is
very good for $\Phi \leq 0.1$.  Figure~\ref{fig:Ra-crit-equal-phi}
shows the variation of $\Ray_c$ and $k_c$ as function of $\Phi$ as
computed by the Chebyshev-collocation approach (in solid symbols) as
well as the analytical value classically obtained for non-penetrating
conditions and the small $\Phi$ expansion. Additionally,
figure~\ref{fig:PhiModeMax} shows the variation of the maximum of
profiles of $P$, $U$ and $W$, that of $\Theta$ being set to 1, as well as the difference between the critical
Rayleigh number for uniform translation ($24\Phi$) and that for a
deforming mode, each as function of $\Phi$. It shows the consistency
between the calculations using the Chebyshev-collocation approach and
the low $\Phi$ development.

At low $\Phi$, the wavelength of the first unstable mode
  tends to infinity as $\sim1/\sqrt{\Phi}$, which means that
  deformation of the solid becomes negligible. Accordingly, the
  viscous stress ceases to be a limiting factor for the flow and
  $Ra_c/\Phi$, which contains no viscosity, tends to a constant
  value. This ratio,
\begin{equation}
  \label{eq:Ra-phi}
  \frac{\Ray}{\Phi}=\frac{\rho \alpha \Delta T d^2}{\Delta \rho^\pm
  \kappa \tau_\phi} \equiv \frac{\Delta\rho_T}{\Delta \rho^\pm}\frac{\tau_\kappa}{\tau_\phi},
\end{equation}
is the ratio of the driving thermal density difference
  $\Delta\rho_T$ to that involved in the phase change, times the ratio
  of the thermal timescale to the phase change one, and can be
  considered as the effective Rayleigh number in the low $\Phi$
  limit.

\begin{figure}
  \centering
  \includegraphics[width=\textwidth]{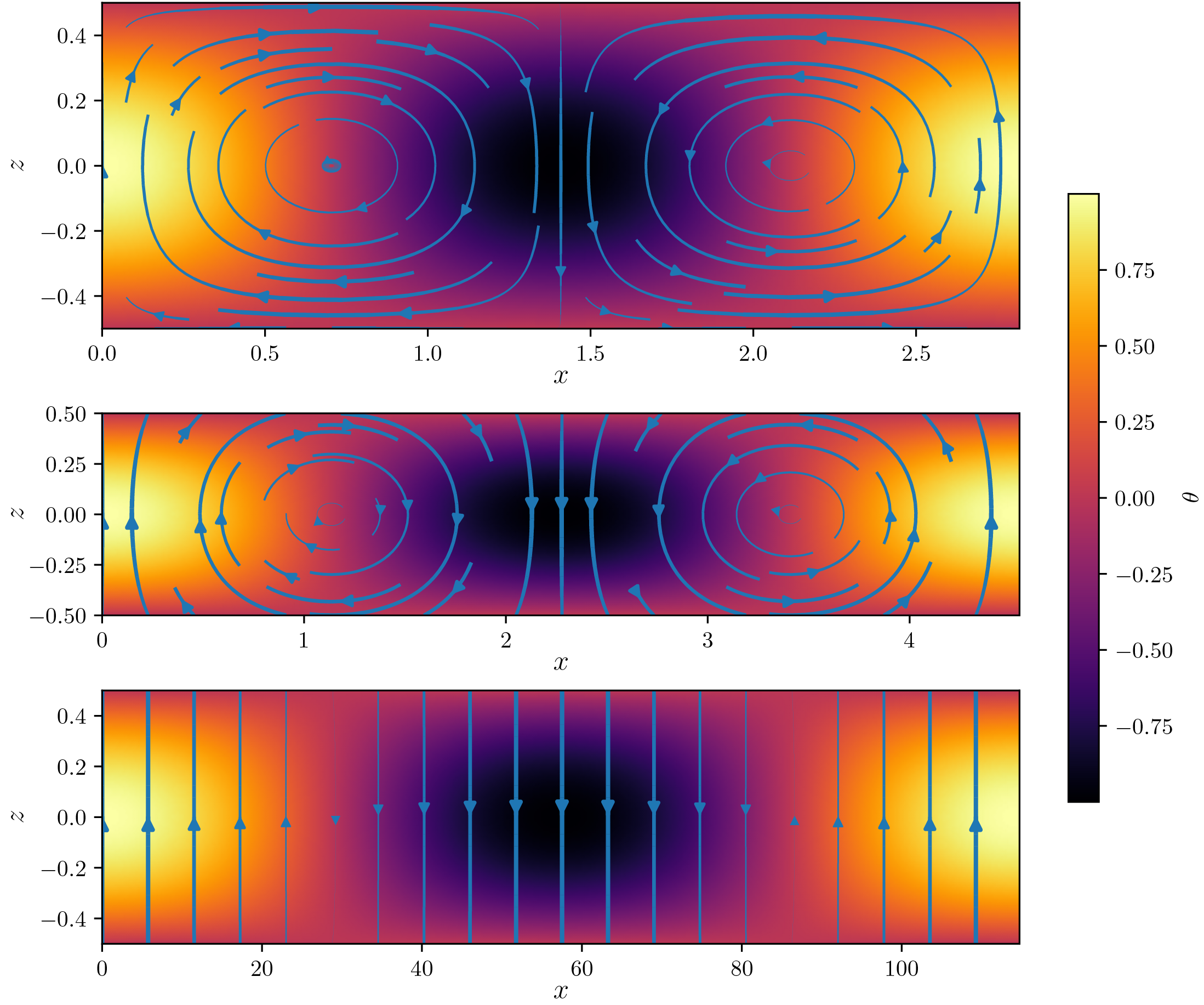}
  \caption{First unstable mode for three different values of 
    $\Phi^+=\Phi^-$: $10^5$ (top), $10$ (middle) and $10^{-2}$ (bottom). The color
    represents temperature and the flow lines thickness is
    proportional to the norm of the velocity. Note that for the bottom
    panel the axis are scaled differently owing to the large
    wavelength of the mode.}
  \label{fig:modes}
\end{figure}

Figure~\ref{fig:modes} shows the first unstable mode for different
values of the phase change parameter. In the case of $\Phi=10^5$, the
critical Rayleigh number and wavenumber are very close to that
obtained using classical non-penetrating boundary conditions
(fig.~\ref{fig:Ra-crit-equal-phi}) and so is the first unstable
mode. For $\Phi=10$, the critical Rayleigh number has already
decreased significantly ($\Ray_c=190$), the critical wavelength
significantly increased ($\lambda_c=4.55$) and the critical mode displays streamlines that
cross both boundaries. For $\Phi=10^{-2}$, the critical Rayleigh
number is a bit less than $0.24$, the critical wavelength is about
$115$ and streamlines are essentially vertical. At each horizontal position,
this mode of convection has exactly the same shape as the linearly
unstable translation mode but it is modulated laterally, with a very
long wavelength that increases as $\sim 1/\sqrt{\Phi}$ when
$\Phi\rightarrow 0$. The fact that this makes the critical Rayleigh
number smaller than that for pure solid-body translation is rather
mysterious. 

\begin{figure}
  \centering
  \includegraphics[width=0.8\textwidth]{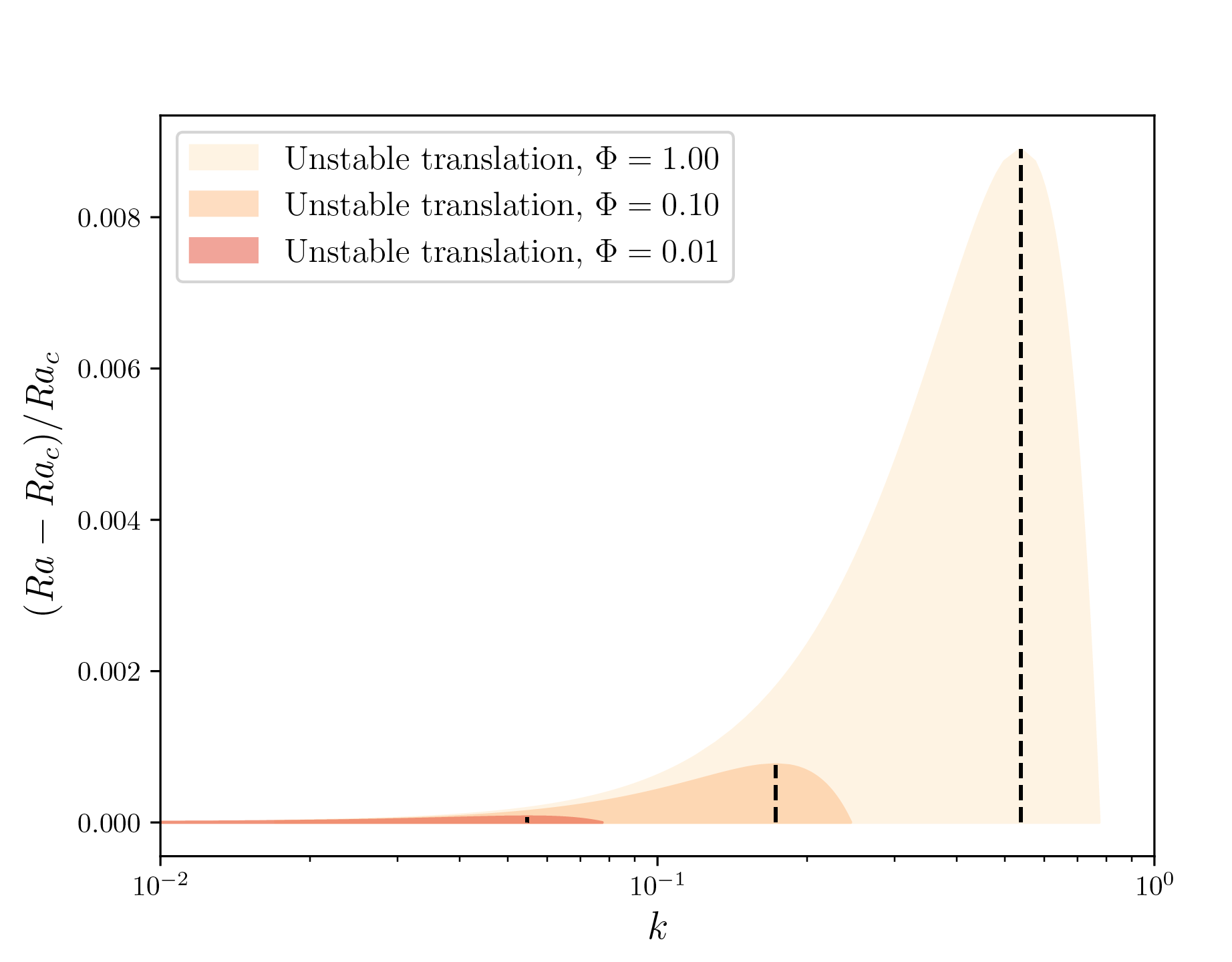}
  \caption{Range of wave numbers as function of the reduced Rayleigh
    number for which the translation solution is unstable versus
    deforming modes. Three different shaded regions for three
    different values of $\Phi$ are represented. For each shaded area,
    the dashed line represents the values of the wave number giving the
    maximum growth rate as function of the reduced Rayleigh number.}
  \label{fig:kx-Ra-instab}
\end{figure}

\begin{figure}
  \centering
  \includegraphics[width=0.8\textwidth]{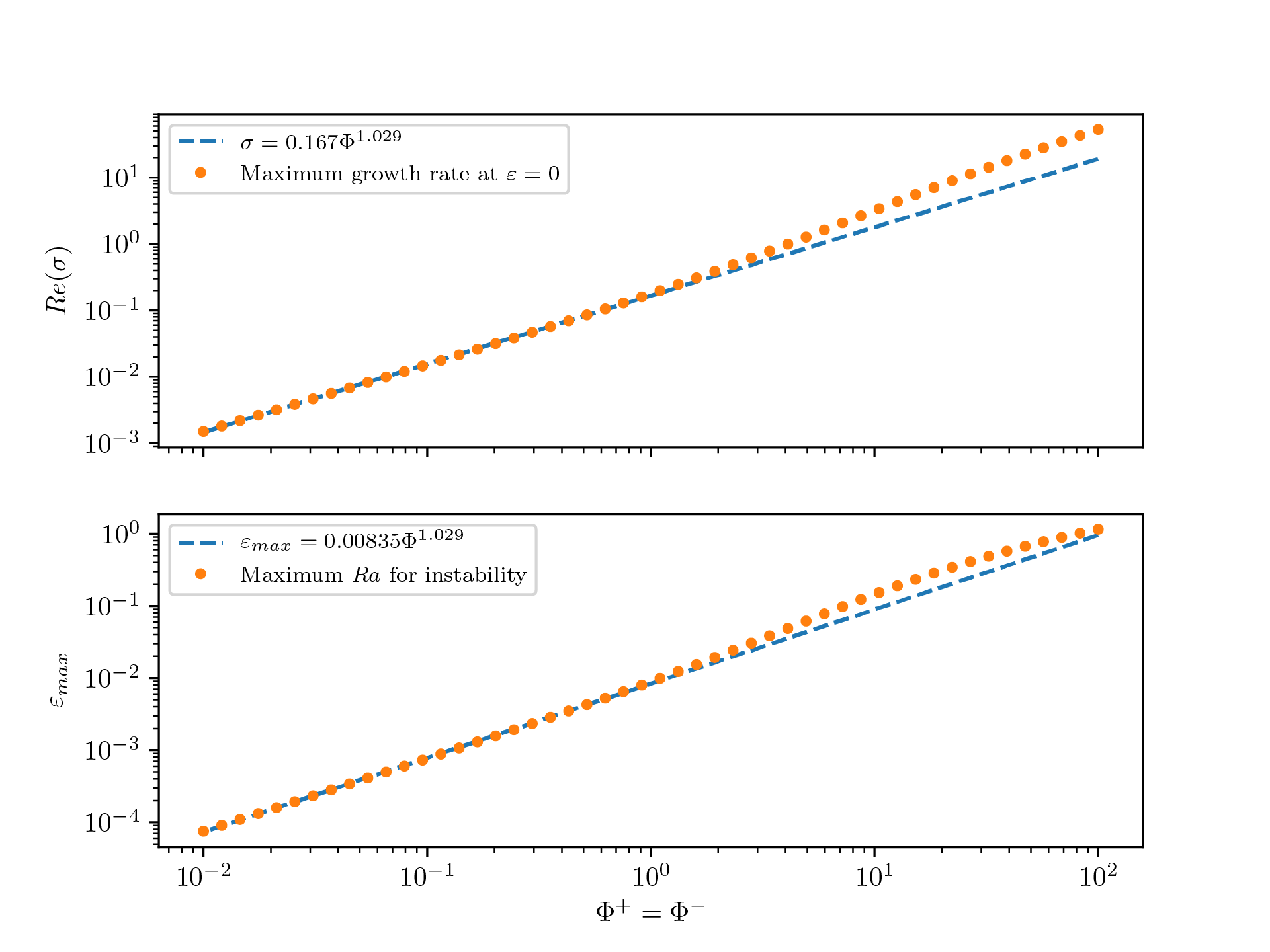}
  \caption{Maximum growth rate for a non-null $k$ mode at the
    critical Rayleigh number for the onset of the translation mode
    (top) and maximum reduced Rayleigh number for a positive growth
    rate of a deforming instability over a finite amplitude
    translation mode (bottom), as function of the phase change number.}
  \label{fig:phi-ramax-sigmax}
\end{figure}

The critical Rayleigh number for the instability for the non-null $k$
mode is always lower than that for pure translation, as shown by
Eq.~\eqref{eq:Ray-phi} and fig.~\ref{fig:Ra-crit-equal-phi} and should therefore always be favored. This
might be true in an infinite layer but, in practical cases, the
horizontal direction is periodic, either in numerical models or in a
planetary mantle. In that case, the minimal value of $k$ that can be
attained is $2\upi/L$ with $L$ the horizontal periodicity. If the
value of $k$ corresponding to the critical Rayleigh number is smaller
than $2\upi/L$, the translation mode could still be favored. 
\begin{figure}
  \centering
  \includegraphics[width=0.8\textwidth]{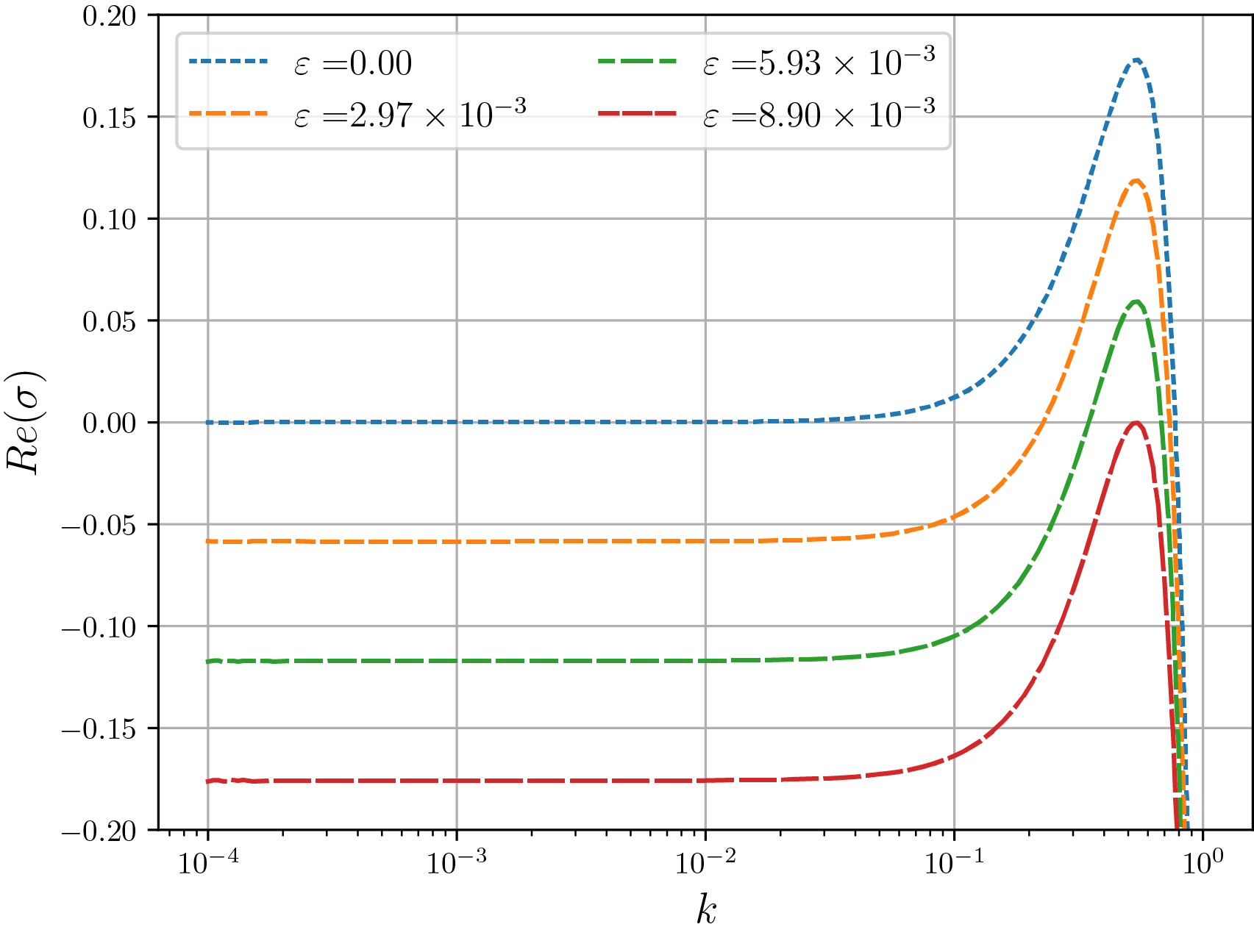}
  \caption{Growth rate of deforming perturbation over a steady
    translating solution as function of the perturbation wavenumber $k$,
    for different values of the reduced Rayleigh
    number $\varepsilon=(\Ray-\Ray_c)/\Ray_c$ and for
    $\Phi^+=\Phi^-=1$.}
  \label{fig:sigma_Ra}
\end{figure}
The study of the stability of the uniformly translating solution with
respect to laterally varying modes is a simple extension to the
stability of the conduction solution. Considering now that $(p,
\sbf{v}, \theta)$ are infinitesimal perturbations with respect to the
steady translation solution $(p_t, w_t\hat{\sbf{z}}, T_t)$, the only
equation to be modified compared to that treated in
section~\ref{sec:linear-stability} at infinite Prandtl number is the
temperature equation that now reads
\begin{equation}
  \label{eq:T-stab-trans}
  \left(\Dif^2-k^2\right) \Theta -w_t \Dif{\Theta} - W \Dif{T_t} =
  \sigma\Theta
\end{equation}
instead of equation~\eqref{eq:energy-linear}. Using the steady
translation solution provided in
section~\ref{sec:steady-state-transl}, this equation can be
implemented in the stability calculation to compute the growth rate of
a deforming perturbation of wavenumber $k$ when a steady translation solution is in
place for a given Rayleigh number above the critical value for the
translation solution. We denote by $\varepsilon=(\Ray-\Ray_c)/\Ray_c$ the
reduced Rayleigh number, $\Ray_c=12(\Phi^++\Phi^-)$ being here the
critical value for the onset of uniform translation. When $\varepsilon$ tends to zero, the
translation velocity $w_t$ tends to zero and the system of equations tends
to that solved for the stability of the steady conduction
solution. But since $\varepsilon=0$ corresponds to the critical
Rayleigh number for the translation solution that is finitely greater
than the critical value for the instability with finite $k$, we
expect a finite instability growth rate in a finite band of wave
numbers. We therefore expect an infinitely slow translation solution
to be unstable with respect to deforming modes. However, when the
Rayleigh number is increased above the critical value for the
translation mode, we expect this translation mode with a finite velocity to become
more stable since perturbations with a finite $k$ are then transported
away by translation. Figure~\ref{fig:sigma_Ra} indeed shows that, for a given
value of the phase change number $\Phi$ (equal for both boundaries here), increasing the Rayleigh number
above the critical value for the translation mode, and therefore the
steady state translation velocity, the linear growth rate of
the deforming mode decreases. For a given Rayleigh number, the growth
rate curve as function of wave number displays a maximum and this
maximum decreases with Rayleigh number and eventually becomes
negative. There is therefore a maximum Rayleigh number beyond which
the translation solution is linearly stable against any deforming
perturbation. Figure~\ref{fig:kx-Ra-instab} shows the range of
unstable modes in the $k-\varepsilon$ space for three different
values of the phase change number. The range of Rayleigh numbers above
the critical one for translation that allows the finite $k$
instabilities to develop shrinks when $\Phi$ decreases and the
translation mode becomes increasingly more
relevant. Figure~\ref{fig:phi-ramax-sigmax} shows that the maximum
growth rate of the instability at $\varepsilon=0$ varies linearly with
$\Phi$ and so does the maximum value of $\varepsilon$ for an
instability to develop. The wave number for the instability is found
to be equal to that for the instability of the conductive solution (fig.~\ref{fig:kx-Ra-instab}) and
therefore varies as $\sqrt{\Phi}$ (fig.~\ref{fig:Ra-crit-equal-phi}).

\subsection{Weakly non-linear analysis}
\label{sec:weakly-non-linear}
Going beyond the linear stability is necessary to assess the behaviour
of the system at Rayleigh numbers larger than the critical value, in
particular to investigate the heat transfer efficiency of the
convective system. We here follow the approach classically developed for weakly nonlinear
dynamics \citep{Malkus_Veronis58,Schluter_etal65,Manneville2004}. The system of partial differential
equations~\eqref{eq:momentum}-~\eqref{eq:energy} is separated into its
linear and nonlinear parts as
\begin{equation}
  \label{eq:lin-nonlin}
  \mbf{L}(\partial_t,\partial_x, \partial_z,\Ray) \mbf{X}=\mbf{N}(\mbf{X},\mbf{X}),
\end{equation}
with $\mbf{X}=(p; u; w; \theta)^T$ and for an infinite Prandtl
case
\begin{equation}
  \label{eq:nonLinear}
  \mbf{L} = \left(
    \begin{array}{cccc}
      0 & \partial_x & \partial_z & 0\\
      -\partial_x & \nabla^2 & 0 & 0\\
      -\partial_z & 0 & \nabla^2 & \Ray \\
      0 & 0 & 1 & \nabla^2-\partial_t
    \end{array}
\right),\quad 
\mbf{N}(\mbf{X}_l,\mbf{X}_m)=\left[
    \begin{array}{c}
      0\\
      0\\
      0\\
      u_l\partial_x\theta_m + w_l\partial_z\theta_m
    \end{array}
  \right]. 
\end{equation}
The linear operator is further developed around the critical Rayleigh
number as
\begin{equation}
  \label{eq:L-M}
  \mbf{L}=\mbf{L}_c-(\Ray-\Ray_c)\mbf{M}.
\end{equation}
By giving $\Ray_c$ as weight to the $\theta$ part in the dot product $\braket{\bullet|\bullet}$, it
can be shown that the operator $\mbf{L}_c$ is self-adjoint (Hermitian),
$\braket{\mbf{X}_2|\mbf{L} \mbf{X}_1}=\braket{\mbf{L}\mbf{X}_2|\mbf{X}_1}$
(see appendix for details). Among other things, it implies that all its
eigenvalues are real and the marginal state is characterized by
$\partial_t=0$. 
The solution $\mbf{X}$ and the Rayleigh number are developed as
\begin{align}
  \label{eq:X-dev}
  \mbf{X} & = \epsilon \mbf{X}_1
            +\epsilon^2\mbf{X}_2+\epsilon^3\mbf{X}_3+...\\
  \label{eq:Ra-dev}
  \Ray & = \Ray_c + \epsilon\Ray_1 +\epsilon^2\Ray_2+... 
\end{align}
and equation~\eqref{eq:lin-nonlin} leads to a set of equations for the
increasing order of $\epsilon$:
\begin{align}
  \label{eq:order1}
  \mbf{L}_c\mbf{X}_1& = \mbf{0},\\
  \label{eq:order2}
  \mbf{L}_c\mbf{X}_2& = \mbf{N}(\mbf{X}_1, \mbf{X}_1) + \Ray_1 \mbf{M} \mbf{X}_1,\\
  \label{eq:order3}
  \mbf{L}_c\mbf{X}_3& = \mbf{N}(\mbf{X}_1, \mbf{X}_2) +
                      \mbf{N}(\mbf{X}_2, \mbf{X}_1) 
                      + \Ray_1 \mbf{M} \mbf{X}_2
                      + \Ray_2 \mbf{M} \mbf{X}_1,\\
  \label{eq:order_n}
  \mbf{L}_c\mbf{X}_n& = \sum_{l=1}^{n-1} \mbf{N}(\mbf{X}_l,
                      \mbf{X}_{n-l})
                      + \sum_{l=1}^{n-1}\Ray_l \mbf{M} \mbf{X}_{n-l}.
\end{align}
Equation~\eqref{eq:order1} is simply that of the linear stability
problem and its solution is $\mbf{X}_1=\mbf{X}_c$ which can be suitably
normalised such that the maximum value of W is 1. 
Taking the scalar product of equations of subsequent orders by
$\mbf{X}_1$ and making use of the Hermitian properties of $\mbf{L}_c$ 
provides solvability conditions (Fredholm alternative)
that determine the values of $\Ray_i$. For $\Ray_1$ one gets:
\begin{equation}
  \label{eq:Ray-1}
  \Ray_1=-\frac{\braket{\mbf{X}_1|\mbf{N}(\mbf{X}_1,\mbf{X}_1)}} 
  {\braket{\mbf{X}_1|\mbf{M}\mbf{X}_1}}.
\end{equation}
The $x$ dependence of $\mbf{X}_1$ is of the form $\ee^{\ii k_c x}$, i.e.
\begin{equation}
  \label{eq:Xc-kc}
  \mbf{X}_1 = \mbf{Z}_{1,1}(z) \ee^{\ii k_c x} + c.c.,
\end{equation}
with $\mbf{Z}_{1,1}(z)=(P_{1,1}(z);
  U_{1,1}(z); W_{1,1}(z); \Theta_{1,1}(z))^T$ the vector composed of
  the four vertical modes for all four variables, at degree 1 of 
  weakly non-linear development (first index) and first mode in the
  horizontal direction (second index).

Then, $\mbf{N}(\mbf{X}_1,\mbf{X}_1)$ contains two contributions to
its $x$ dependence, one constant and one in $\ee^{\ii 2k_c x}$. It is therefore orthogonal to
$\mbf{X}_1$ and it can then be concluded that $\Ray_1=0$.
The general solution to equation~\eqref{eq:order2} is the sum of the
solution to the homogeneous equation and a particular solution of the
equation with a right-hand-side. Since we are seeking a solution $\mbf{X}_2$ which
adds to $\mbf{X}_1$, i.e. orthogonal to it, and since $\mbf{X}_1$ is the general solution to the
homogeneous equation, only the particular solution is of interest. The
$x$ dependence of $\mbf{X}_2$ will contain a constant value
of the form $\mbf{Z}_{2,0}(z)$
and a term of the form $\mbf{Z}_{2,2}(z)\ee^{\ii 2k_c x}$. 
Computing the scalar product of
equation~\eqref{eq:order3} by ${\mbf{X}}_1$ gives the value of
$\Ray_2$:
\begin{equation}
  \label{eq:Ray-2}
  \Ray_2=-\frac{\braket{\mbf{X}_1|\mbf{N}(\mbf{X}_2,\mbf{X}_1)}
  +\braket{\mbf{X}_1|\mbf{N}(\mbf{X}_1,\mbf{X}_2)}}
  {\braket{\mbf{X}_1|\mbf{M}\mbf{X}_1}}.
\end{equation}
$\mbf{X_2}$ containing a term proportional to $\mathrm{e}^{i 2k_c x}$
and a term independent of $x$, $\mbf{N}(\mbf{X}_2,\mbf{X}_1)$ and $\mbf{N}(\mbf{X}_1,\mbf{X}_2)$ have contributions of the
form $\mathrm{e}^{\pm i k_c x}$ which can resonate with $\mbf{X}_1$
and make $\Ray_2$ non-null. In that case, the amplitude parameter is, to leading order,
\begin{equation}
  \label{eq:amplitude}
  \epsilon = \sqrt{\frac{\Ray-\Ray_c}{\Ray_2}}.
\end{equation}
The procedure can be extended to any higher
order and the general behaviour can be predicted by recursive
reasoning. In particular, it is easy to show that solutions of even and
odd order contain contributions to their $x$ dependence as even and
odd powers of $\ee^{\ii k_c x}$ up to their order value, i.e.
\begin{align}
  \label{eq:even-order}
  \mbf{X}_{2n} & = \sum_{l=0}^n\mbf{Z}_{2n,2l}(z)\ee^{\ii 2l k_c x} + c.c.,\\
  \label{eq:odd-order}
  \mbf{X}_{2n+1} & = \sum_{l=0}^n\mbf{Z}_{2n+1,2l+1}(z)\ee^{\ii (2l+1) k_c x} + c.c.,
\end{align}
the vertical normal mode $\mbf{Z}_{n,l}=(P_{n,l}(z);
  U_{n,l}(z); W_{n,l}(z); \Theta_{n,l}(z))^T$ being indexed with the order
$n$ of the solution and harmonic number $l$ in the $x$ dependence. 
It also appears recursively that
\begin{align}
  \label{eq:Ra2n}
  \Ray_{2n} & = -
              \frac{\sum_{l=1}^{2n}\braket{\mbf{X}_1|\mbf{N}(\mbf{X}_l,\mbf{X}_{2n+1-l})}
+\sum_{l=1}^{n-1}\Ray_{2l}\braket{\mbf{X}_1|\mbf{M}\mbf{X}_{2(n-l)+1}}}
              {\braket{\mbf{X}_1|\mbf{M}\mbf{X}_1}},\\
  \label{eq:Ra2n1}
  \Ray_{2n+1} & = 0.
\end{align}
This is true for orders 1 and 2, as explained above and, assuming it
holds up to degrees $2n-1$ and $2n$, the expressions for degrees
$2n+1$ and $2n+2$ can be predicted from
equation~\eqref{eq:order_n}. First, equation~\eqref{eq:order_n} of
order $2n+1$ includes on the right-hand-side only terms up to degree
$2n$ and can be used to predict the form of $\mbf{X}_{2n+1}$. Each
term of the form $\mbf{N}(\mbf{X}_l,\mbf{X}_{2n+1-l})$ contains only
odd powers of $\ee^{\ii k_c x}$ since it is composed of products of
even (resp. odd) and odd (resp. even) polynomials of $\ee^{\ii k_c x}$
for $l$ even (resp. odd). Each term of the form
$\Ray_l\mbf{M}\mbf{X}_{2n+1-l}$ is either null for $l$ odd or an odd
polynomial of $\ee^{\ii k_c x}$ for $l$ even. Summing up, the
right-hand-side of the equation being an odd polynomial of $\ee^{\ii
  k_c x}$, the solution to the equation is of the form~\eqref{eq:odd-order}.

Taking the dot product of
equation~\eqref{eq:order_n} of order $2n+2$ by $\mbf{X}_1$
and using the Hermitian character of $\mbf{L}_c$ provides the equation
for $\Ray_{2n+1}$. Starting first with the last term on the
right-hand-side, all the terms in the sum except the one in
$\Ray_{2n+1}$ drop out either because $\Ray_l$ is null for $l$ odd or
because the dot product $\braket{\mbf{X}_1|\mbf{M}\mbf{X}_{2n+2-l}}=0$
for $l$ even since $\mbf{X}_{2n+2-l}$ then contains only even powers of
$\ee^{\ii k_c x}$. We are left with
$\Ray_{2n+1}\braket{\mbf{X}_1|\mbf{M}\mbf{X}_1}$. Considering the
first sum on the right-hand-side, each term
$\mbf{N}(\mbf{X}_l,\mbf{X}_{2n+2-l})$ is an even polynomial of
$\ee^{\ii k_c x}$, as the product of either two even polynomials (for
$l$ even) or two odd polynomials (for $l$ odd). Therefore, each of
these terms is orthogonal to $\mbf{X}_1$ and $\Ray_{2n+1}=0$. The same
equation~\eqref{eq:order_n}$_{2n+2}$ contains only even powers of
$\ee^{\ii k_c x}$ on the right-hand-side and this justifies
equation~\eqref{eq:even-order} for the order $2n+2$. 

Finally, equation~\eqref{eq:Ra2n}$_{2n+2}$ is obtained by simply taking
the dot product of equation~\eqref{eq:order_n}$_{2n+3}$ by $\mbf{X}_1$. 

An important diagnostic for convection is the heat transfer
efficiency measured by the dimensionless mean heat flux density, the
Nusselt number $\Nu$. Since the temperature is uniform on each
horizontal boundary and the average vertical velocity is null for the
deforming mode considered here, the advective heat transfer across the
horizontal boundaries is null. Therefore, the Nusselt number can
easily be computed by taking the vertical derivative of the
temperature at either boundary. In the Fourier decomposition used for
the non-linear analysis, only the zeroth order term in $\ee^{\ii k_c
  x}$ contribute to the horizontal average and they only appear in
terms that are even in the $\epsilon$ development. Restricting ourselves here
to an order 2 development, the Nusselt number can be computed as
\begin{equation}
  \label{eq:nusselt-gen}
  \Nu = 1 - \epsilon^2 D \theta_{2,0}\left(\frac{1}{2}\right) = 
  1 - D \theta_{2,0}\left(\frac{1}{2}\right) \frac{\Ray_c}{\Ray_2}\frac{\Ray-\Ray_c}{\Ray_c}
\end{equation}
where equation~\eqref{eq:amplitude} was used. This equation shows the
classical result that the convective heat flow, $\Nu -1$, increases
linearly with the reduced Rayleigh number
$\varepsilon=(\Ray-\Ray_c)/\Ray_c$ for small values of $\varepsilon$ and the determination of the
coefficient of proportionality, $A$, is the main goal of the weakly
non-linear analysis presented here. Note that
$\mbf{N}(\mbf{X}_2,\mbf{X}_1)$ and $\mbf{N}(\mbf{X}_1,\mbf{X}_2)$ only
have a non-zero component only along the $\theta$ space
(eq.~\ref{eq:nonLinear}) so that, because of our definition of the
dot product (\S~\ref{sec:self-adj}) and using
equation~\eqref{eq:Ray-2}, $\Ray_2$ is proportional to $\Ray_c$. 

The procedure just outlined can be applied to the case with classical
boundary conditions. In particular, for free-slip non-penetrating boundary conditions,
the problem can be solved analytically
\citep{Malkus_Veronis58,Manneville2004}. Starting with the vertical
velocity in the critical mode as $w_1=\sin k x \cos \pi z$, one gets
$\theta_1=\left(\pi^2+k^2\right)^{-1} \sin k x \cos \pi z$,
$\Ray_c=\left(\pi^2+k^2\right)^3/k^2$, $w_2=0$,
$\theta_2=\left(8 \pi \left(\pi^2+k^2\right)\right)^{-1} \sin 2 \pi z$
 and
$\Ray_2=\left(\pi^2+k^2\right)^2/8 k^2$. This gives $A = - D
\theta_{2,0}\left(1/2\right) \Ray_c/\Ray_2=2$. 

\begin{figure}
  \centering
  \includegraphics[width=\textwidth]{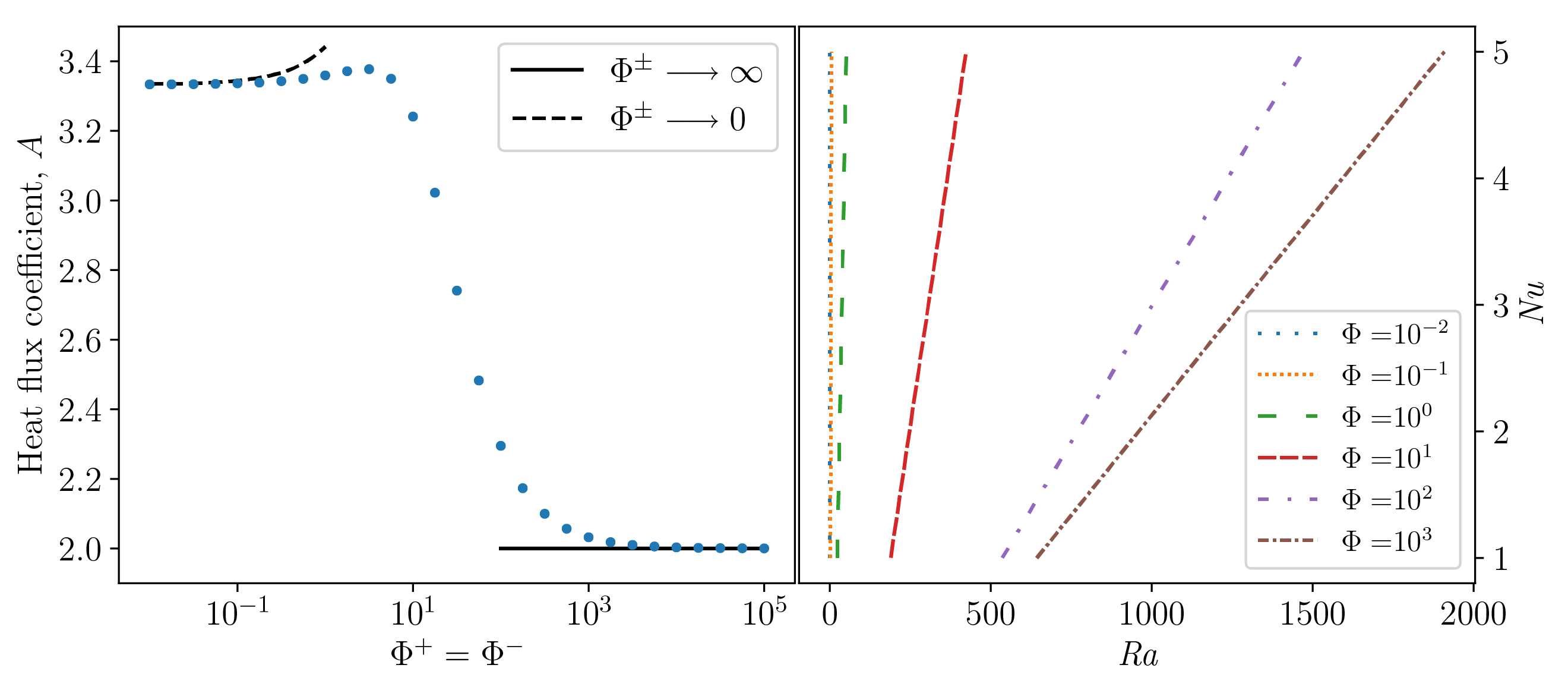}
  \caption{Heat flux coefficient as a function of the phase change
    numbers, equal to each other (left), and Nusselt number as a
    function of Rayleigh number for different values of $\Phi^\pm$ (right). The solid line gives the limit of two
    non-penetrating boundaries while the dashed line represents the
    first order development obtained for $\Phi \rightarrow 0$ (eq.~\ref{eq:HF-coeff}).}
  \label{fig:HF-BotTop}
\end{figure}
Similarly, the low $\Phi$ expansion of the linear mode,
equations~\eqref{eq:kc-phi}--~\eqref{eq:P-lead}, can be used to
compute the behaviour of coefficient $A$ at low $\Phi$ values.
We choose $\Theta_{max}=1/16$ to have a normalisation consistent with
the one above\footnote{The amplitude of $\mbf{X}_1$ is not defined
by the linear problem and changing its normalisation, say by
multiplying it by a factor $a$, leads to $\mbf{X}_2$ and $\Ray_2$
multiplied by $a^2$, so that by virtue of
equation~\eqref{eq:amplitude}, the total solution $\mbf{X}$ is
unchanged.} and the solution at order 2 is searched in the form of
$z$ polynomials, and we get, to order 1 in $\Phi$,
\begin{align}
  \label{eq:theta-2}
  \theta_2 & = -\frac{z}{48}\left(z^2-\frac{1}{4}\right)
             \left[1 +\left( 1-\frac{\Phi}{64} \right)  \cos{2 k_c x} \right],\\
  \label{eq:U-2}
  u_2 & = - \frac{\sqrt{\Phi}}{192\sqrt{2}} \sin{2 k_c x},\\
  \label{eq:W-2}
  w_2 & = \frac{z \Phi}{256} \cos{2 k_c x},\\
  \label{eq:ra2}
  \Ray_2 & = \frac{1}{320} - \frac{43 \Phi}{430080}.
\end{align}
The heat flux coefficient is then, to order 1 in $\Phi$:
\begin{equation}
  \label{eq:HF-coeff}
  A = \frac{4480}{1344 - 43 \Phi}.
\end{equation}

Figure~\ref{fig:HF-BotTop} represents the value of the heat flux
coefficient $A$ as function of $\Phi$ obtained using the
Chebyshev-collocation approach described above (solid circles, see
appendix~\ref{sec:non-lin-term} for details on the calculation of
non-linear terms)
and the two limiting cases of $\Phi\rightarrow\infty$ (solid line) and
$\Phi\rightarrow 0$ (dashed line), which shows a good match.

The heat flux coefficient $A$, which equals 2 for classical
non-penetrating boundaries, tends to $10/3$ when $\Phi\rightarrow
0$. This $\sim50\%$ increase makes the Nusselt number increase when
$\Phi$ tends to zero but the main effect comes from the decrease of
the critical Rayleigh number as $\sim 24 \Phi$, which makes the slope
$\dif{\Nu} / \dif{\Ray}$ go to infinity as $\sim 5/ 36 \Phi$. This is
illustrated on figure~\ref{fig:HF-BotTop} which shows the $\Nu-\Ray$
relationship derived from this analysis for different values of
$\Phi$. The heat transfer efficiency is greatly increased by
decreasing $\Phi$ for two reasons. Firstly, it makes the critical
Rayleigh number decrease so that convection starts with a lower
Rayleigh number. Secondly, the rate at which the Nusselt number
increases with $\Ray$ above its critical value is also drastically
increased when $\Phi$ is decreased.

\section{Solutions with only one phase change boundary}
\label{sec:solutions-with-only}
Let us now consider the case when only one boundary is a liquid-solid
phase change, the other one being subject to a non-penetrating
condition. With the plane layer geometry considered here, the
situation with the upper boundary a phase change is symmetrical to the
one with a lower boundary a phase change. The latter is considered
here since it applies to the dynamics of the icy shells of some
satellites of giant planets \citep{Cadek_etal2016} and possibly to the Earth mantle for a
large part of its history \citep{Labrosse_etal07}.

\begin{figure}
  \centering
  \includegraphics[width=\textwidth]{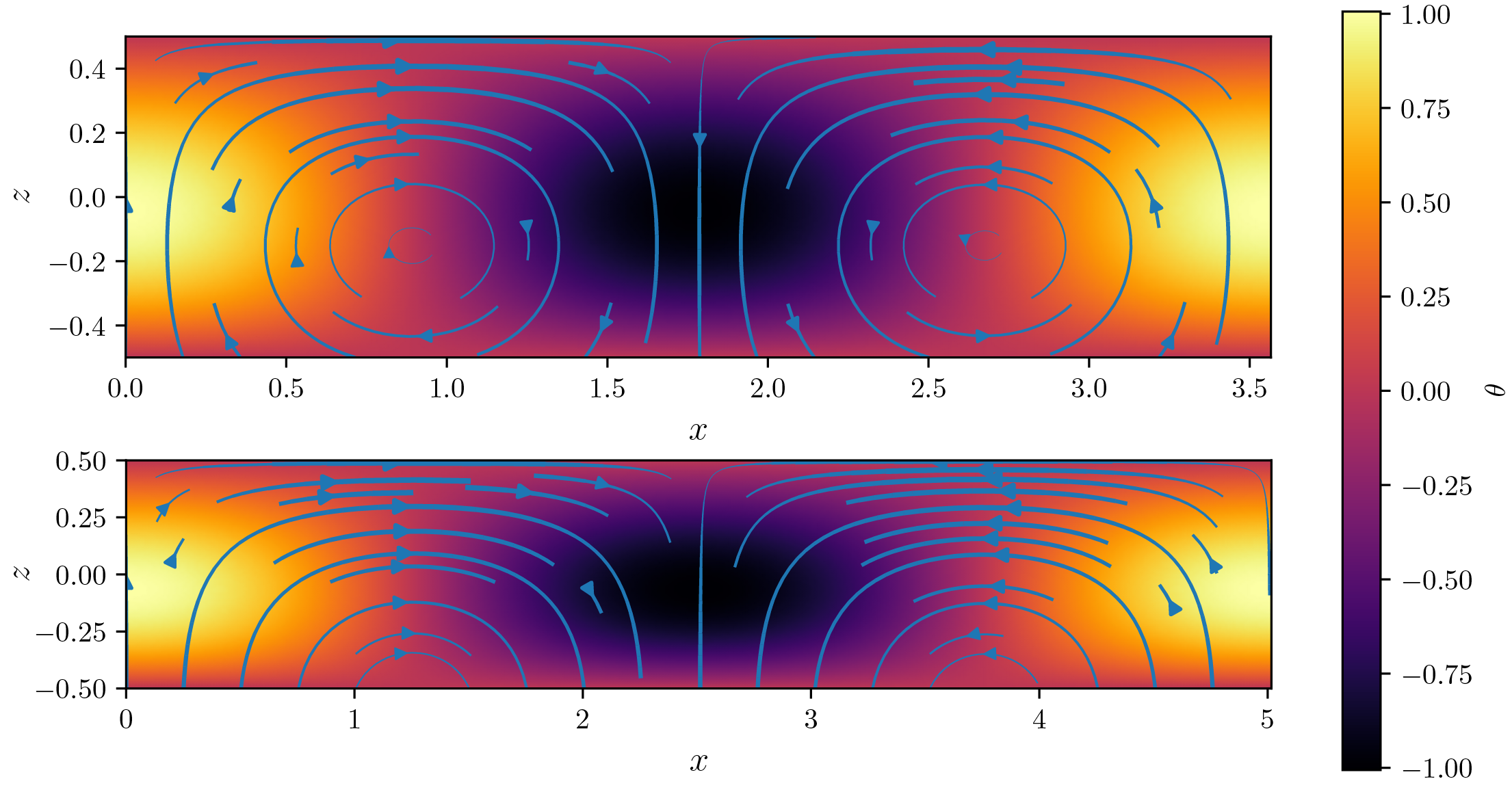}
  \caption{First unstable mode when only the bottom boundary is a
    phase change interface, with $\Phi^-=10$ (top) and
    $\Phi^-=10^{-2}$ (bottom). The temperature anomaly compared to the
    conduction solution is represented in colours and streamlines have
    a thickness proportional to the relative norm of the velocity.}
  \label{fig:openbot}
\end{figure}

The analysis is done in the same way as for the case with a phase
change at both boundaries. Figure~\ref{fig:openbot} shows examples of
the first unstable mode for two different values of $\Phi^-$. The
upper one shows that when $\Phi^-=10$, the convection geometry is not
very different from that with a non-penetrating condition (hereafter
``the classical situation'') but the streamlines are
slightly open at the bottom. The horizontal wavelength at onset,
$\lambda_c=3.57$, is larger than the one for the classical situation
($\lambda_c=2\sqrt{2}$) and the critical Rayleigh number is smaller
($\Ray_c=352$). For $\Phi^-=10^{-2}$, the streamlines are
almost normal to the bottom boundary and the wavelength $\lambda_c=5$ is
about twice the classical one, as if the solution was the upper half
of a classical convective domain. However, the boundary condition
imposed for temperature at the bottom is different from what would be
obtained in that case and the critical Rayleigh number,
$\Ray_c=153$ is about a quarter of the classical one. This can be
understood in a heuristic way: The Rayleigh number can be written as
\begin{equation}
  \label{eq:Ray-heuristic}
  \Ray = \frac{\tau_\nu \tau_\kappa}{\tau_c^2} = \frac{\alpha \Delta T
    g}{d} \frac{d^2}{\nu} \frac{d^2}{\kappa}, 
\end{equation}
with $\tau_c$ the convective time scale associated with acceleration
due to gravity, $\tau_\nu$ the viscous time scale and $\tau_\kappa$
the thermal diffusion time scale. Compared to the classical situation,
we have the same imposed temperature gradient, hence the same
$\tau_c$. Similarly, diffusion happens on the same vertical length
scale and we have the same $\tau_\kappa$. On the other hand, the
bottom boundary imposes no limit to vertical flow and the viscous
deformation is distributed on vertical distance twice the thickness of
the layer, which increases the effective viscous time scale by a factor
4. Therefore, the Rayleigh number imposed here is equivalent to a
value 4 times larger in the classical situation. 

\begin{figure}
  \centering
  \includegraphics[width=\textwidth]{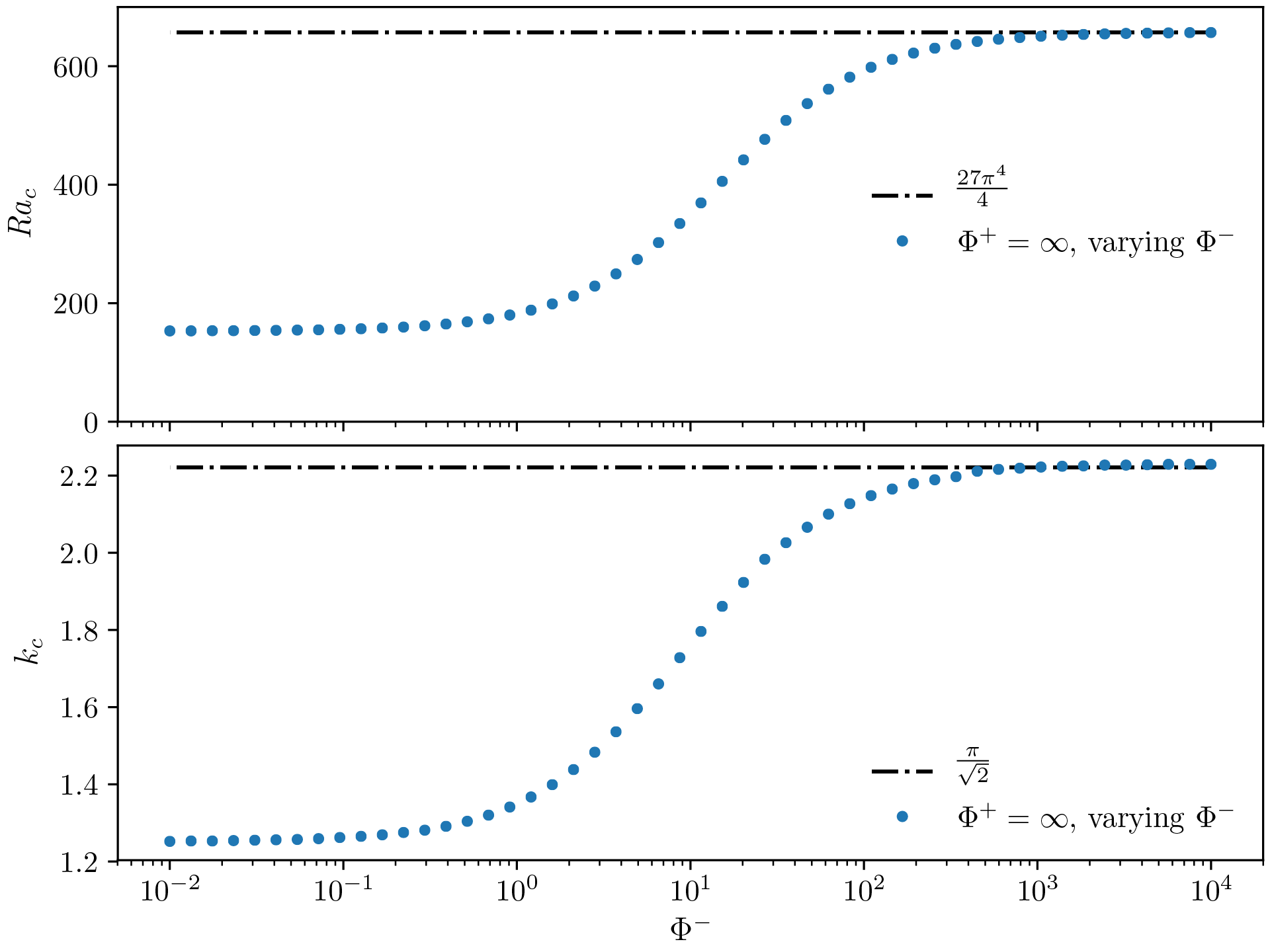}
  \caption{Critical Rayleigh number (top) and wavenumber (bottom) as function
    of the phase change number for the bottom boundary $\Phi^-$, the top one
    having a non-penetrating condition. The dash dotted lines
    represent the classical values obtained for two non-penetrating
    conditions, for reference.}
  \label{fig:Ra-crit-diff-phi}
\end{figure}

Figure~\ref{fig:Ra-crit-diff-phi} shows the variation of the critical
Rayleigh number (top) and wavenumber (bottom) as a function of
$\Phi^-$ and one can see that both tend to a finite value when
$\Phi^-\rightarrow 0$. The mode obtained for $\Phi^-=10^{-2}$ is close
to that limit. Contrary to the situation with a phase change at both
boundaries, the presence of non-penetrating boundary condition implies
that some deformation is always needed for convection to occurs, which
makes viscosity still be a limiting factor at vanishing values of $\Phi^-$.

\begin{figure}
  \centering
  \includegraphics[width=\textwidth]{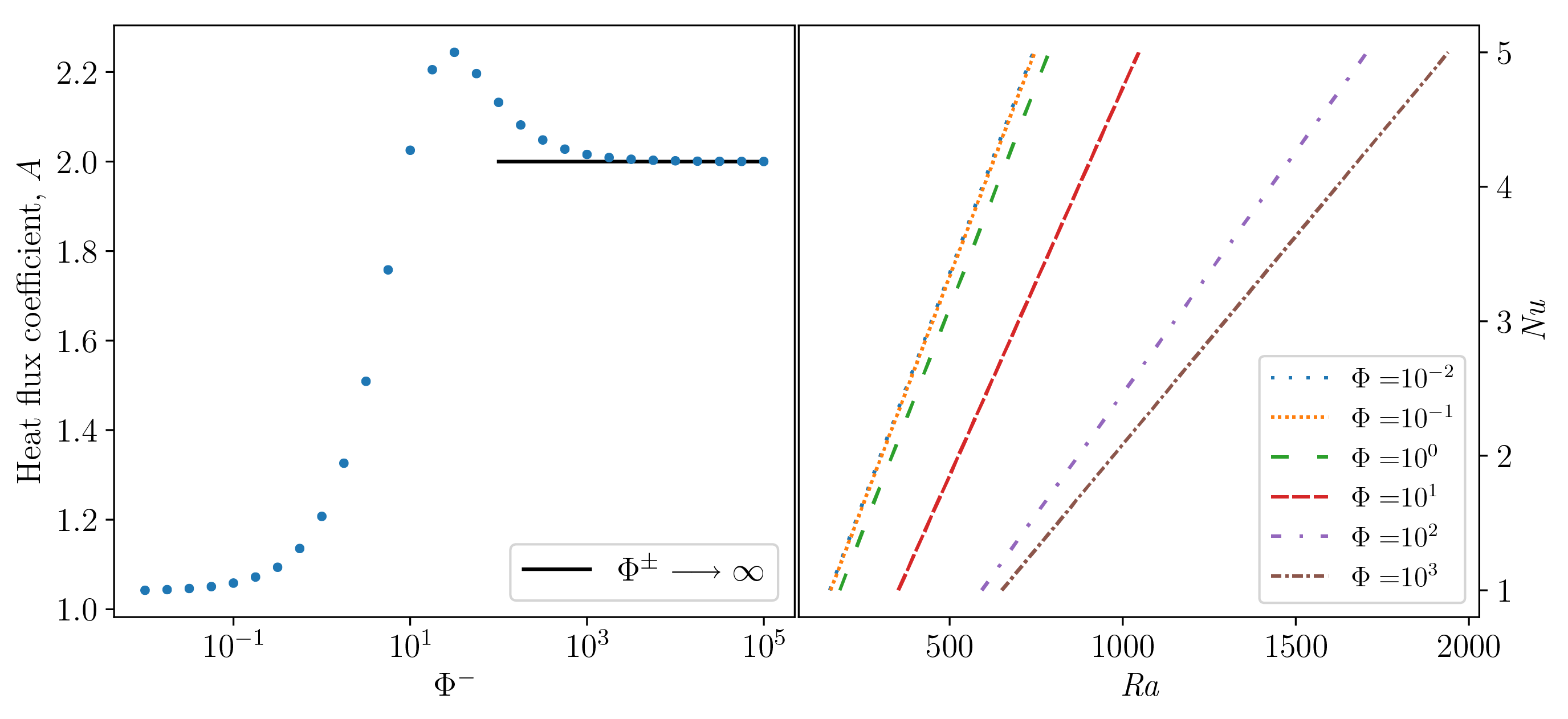}
  \caption{Heat flux coefficient as a function of the bottom phase change
    number $\Phi^-$, the top boundary being non-penetrative (left), and Nusselt number as a
    function of Rayleigh number for different values of $\Phi^-$ (right).}
  \label{fig:HF-BotOnly}
\end{figure}
Considering now the weakly non-linear analysis results,
figure~\ref{fig:HF-BotOnly} shows that the heat flux coefficient for
only one phase change boundary condition tends to a little above 1,
that is about half that for the case for both non-penetrative
boundaries. Combining that with a critical Rayleigh number that is
about four times smaller makes $\dif{\Nu}/\dif{\Ray}$ about twice that
for the classical situation. Therefore, the efficiency of heat
transfer is improved compared to the classical case, both because
convection starts for a smaller Rayleigh number and because the rate
of variation of the Nusselt number with $\Ray$ is about twice larger. 
This is illustrated on the right panel of figure~\ref{fig:HF-BotOnly}.

\begin{figure}
  \centering
  \includegraphics[width=\textwidth]{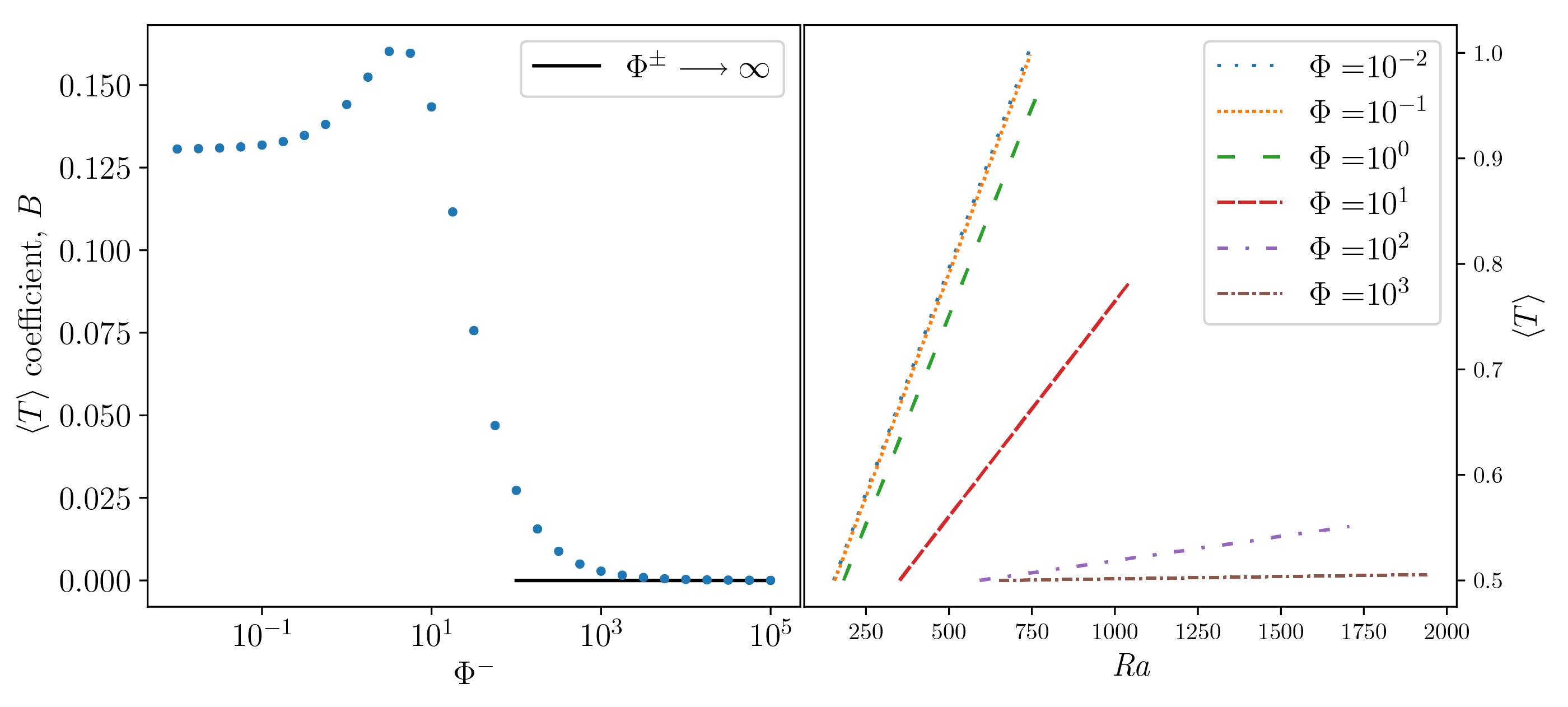}
  \caption{Mean temperature coefficient ($B$ defined in equation~\protect\eqref{eq:meanT}) as
    function of the bottom phase change parameter $\Phi^-$ (left) and
    mean temperature as function of $\Ray$ for different values of
    $\Phi^-$ (right). The range of $\Ray$ values explored is the same as that
    used for figure~\protect\ref{fig:HF-BotOnly}.}
  \label{fig:meant}
\end{figure}
In contrary to the case with both boundaries being a phase change with
equal values of $\Phi$, the case discussed in this section breaks the
symmetry around the $z=0$ plane. In particular, this means that the
mean temperature in the domain is not equal to the average of both
boundaries, $\langle T \rangle \neq 1/2$ in dimensionless form. As for
the Nusselt number (eq.~\ref{eq:nusselt-gen}), a contribution from all even orders in $\epsilon$
is expected, and to the leading order explored here,
\begin{equation}
  \label{eq:meanT}
  \langle T \rangle = \frac{1}{2} + \langle
\theta_{2,0}\rangle \frac{\Ray_c}{\Ray_2}\frac{\Ray-\Ray_c}{\Ray_c}
\equiv  \frac{1}{2} + B \frac{\Ray-\Ray_c}{\Ray_c}.
\end{equation}
The coefficient $B$ defined above is computed exactly for the case of
both non-penetrating boundaries and as expected found to be
null. Figure~\ref{fig:meant} shows the evolution of this coefficient
as function of $\Phi^-$. One can see that it tends to a finite positive
value when in the limit $\Phi^-\rightarrow 0$. Therefore, for small
values of $\Phi^-$, the average temperature is expected to be larger
than $1/2$ (figure~\ref{fig:meant}). For the same range of Rayleigh
number as explored in figure~\ref{fig:HF-BotOnly},
figure~\ref{fig:meant} also shows the evolution of the mean
temperature at the leading order given by
equation~\eqref{eq:meanT}. For low values of $\Phi^-$, the mean
temperature increases rapidly with Rayleigh number. 

\begin{figure}
  \centering
  \includegraphics[width=\textwidth]{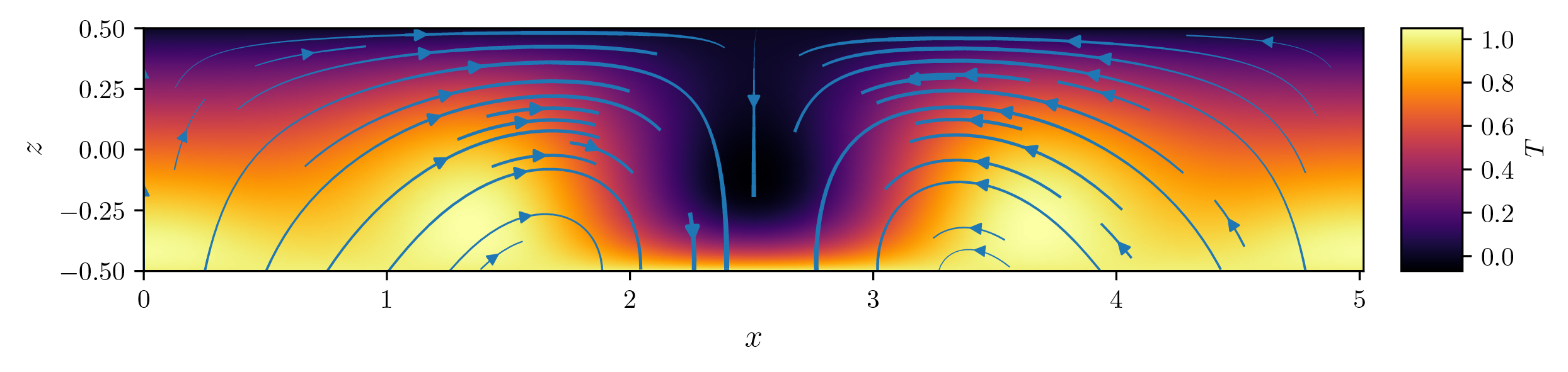}
  \caption{Finite amplitude solution for $\Phi^-=10^{-2}$, $\epsilon=5.58$ and a
    non-penetrating boundary condition at the top. }
  \label{fig:solution-phi1e-2}
\end{figure}
The asymmetry of the mean temperature for low values of $\Phi^-$ is
also expressed in the finite amplitude solution that can be plotted
for a given value of $\epsilon$. The range of validity of such
solutions as function of $\epsilon$ depends on the order of the
development. Computing the solution only up to order 3 in $\epsilon$,
we restrict ourselves to small values of this number and
figure~\ref{fig:solution-phi1e-2} shows the result for $\epsilon=5.58$
corresponding to $\Nu=1.5$. This shows that the down-welling current
is more focused than the up-welling one. This situation is similar to
the case of volumetrically heated convection
\citep[e. g.][]{Parmentier_Sotin2000}, which is not the case
here. Preliminary direct numerical simulations confirm this behaviour
but the full exploration of this question goes beyond the scope of the
present paper. 

\section{Conclusion}
\label{sec:conclusion}
In the context of the dynamics of planetary mantles, convection can
happen in solid shells adjacent to liquid layers. The viscous stress
in the solid builds up a topography of the interface between the
solid and liquid layers. In the absence of mechanisms to erase
topography, its buoyancy equilibrates the viscous stress which
effectively enforces a non-penetrating boundary condition. On the
other hand, if the topography can be suppressed by melting and
freezing at the interface at a faster pace than its building process,
the vertical velocity is not required to be null at the interface. The
non-penetrating boundary condition is then replaced by a relationship
between the normal velocity, its normal gradient, and pressure
(eq.~\ref{eq:BC-wp}) and involving a dimensionless phase change
number, $\Phi$, ratio of the phase change timescale to the viscous
timescale (eq.~\ref{eq:phase-number}). When this number is large, we
recover the classical non-penetrating condition while the limit of low
$\Phi$ authorises a large flow through the boundary.

When both boundaries are characterised by a low $\Phi$ number, a
translating, non-deforming, mode of convection is possible and competes
with a deforming mode with wave number $k$ that decreases as
$\sqrt{\Phi}$, and therefore ressembles translation with alternating
up- and downward direction. The critical Rayleigh number for the onset
of the deforming mode is slightly below that of the translation mode,
$\Ray=24 \Phi$,
but the latter is found to be stable against a deforming instability
when the Rayleigh number is $\sim \Phi^2$ above the critical value. It
is therefore likely to dominate when both boundaries are characterised
by low values of $\Phi$. In both translating and deforming modes of
convection, the heat transfer efficiency, the Nusselt number, is found
to increase strongly with Rayleigh number at small values of $\Phi$.

When only one boundary is a phase change interface with a low value of
$\Phi$, the wavenumber is about half and the critical Rayleigh number is
about a quarter the corresponding values for the classical non-penetrating boundary
condition. Close to onset, a weakly non-linear analysis shows that the
Nusselt number varies linearly with the Rayleigh number with a slope
that is about twice that for both non-penetrating boundary
conditions. The average temperature is also found to increase strongly
with Rayleigh number and the flow geometry is strongly affected, with
down-welling currents more focused than up-welling ones. 

Overall, having the possibility of melting and freezing across one or
both horizontal boundaries of an infinite Prandtl number fluid makes
convection much easier (i.e. the critical Rayleigh number is strongly
reduced), the preferred horizontal wavelength much larger and heat
transfer much stronger, with important potential implications for
planetary dynamics. 

\section{Acknowledgments}
\label{sec:Acknowledgments}
We are thankful to three anonymous reviewers and editor Grae Worster
for comments that pushed us to significantly clarify our paper.
This research has been funded by the french Agence Nationale de la
Recherche under the grant number ANR-15-CE31-0018-01, MaCoMaOc.

\appendix

\section{Self-adjointness of operator $\mbf{L}_c$}
\label{sec:self-adj}
Using a Fourier decomposition for the horizontal decomposition,
$\mbf{L}_c$ simply reads as
\begin{equation}
  \label{eq:Linear}
  \mbf{L}_c = \left(
    \begin{array}{cccc}
      0 & \ii k & \Dif & 0\\
      -\ii k & \Dif^2-k^2 & 0 & 0\\
      -\Dif & 0 & \Dif^2-k^2 & \Ray_c \\
      0 & 0 & 1 &  \Dif^2-k^2
    \end{array}
\right) 
\end{equation}
where the time derivative has been omitted since the linear
instability is found to be stationary. In a linear stability
analysis, adding a growth rate $\sigma$ on the diagonal of the matrix
would not alter the adjoint calculation, as will appear below. The
boundary conditions are given by equations~(\ref{eq:BC-u})
to~(\ref{eq:BC-theta}). In the calculation of the dot product, the
$\theta$ part is given $\Ray_c$ as weight and the horizontal integral
can be factored out:
\begin{equation}
  \label{eq:dot-prod}
  \begin{split}
    \braket{\mbf{X}_2|\mbf{L}_c\mbf{X}_1} = & \int \ee^{\ii (k_2-k_1)}\dif{x}\left[
      \int_{-\frac{1}{2}}^\frac{1}{2}\bar{P}_2\left(\ii k
        U_1+\Dif W_1\right)\dif{z} \right.\\
    &   +\int_{-\frac{1}{2}}^\frac{1}{2}\bar{U}_2\left(-\ii k
      P_1+\left(\Dif^2-k^2\right) U_1\right)\dif{z}\\
    &  +\int_{-\frac{1}{2}}^\frac{1}{2}\bar{W}_2\left(-\Dif
        P_1+\left(\Dif^2-k^2\right) W_1+\Ray_c \Theta_1\right)\dif{z}\\
    &  \left. +\Ray_c \int_{-\frac{1}{2}}^\frac{1}{2}\bar{\Theta}_2 \left(W_1
        +\left(\Dif^2-k^2\right) \Theta_1\right)\dif{z}\right]
  \end{split}
\end{equation}
where the overbar means complex conjugate.
Since the $x$ part poses no difficulty, we only consider the $z$ part,
which we denote as $\braket{\bullet|\bullet}_z$. Reordering the
different integrals in Eq.~\eqref{eq:dot-prod} so that terms of
$\mbf{X}_1$ are factored out and performing integrations by part on
each term including $\Dif$, we get
\begin{equation}
  \label{eq:dot-prod-z}
  \begin{split}
    \braket{\mbf{X}_2|\mbf{L}_c\mbf{X}_1}_z = &  
    \int_{-\frac{1}{2}}^\frac{1}{2}\left(-\ii k \bar{U}_2+\Dif \bar{W}_2\right) P_1\dif{z} \\
    &   +\int_{-\frac{1}{2}}^\frac{1}{2}\left(\ii k
      \bar{P}_2+\left(\Dif^2-k^2\right)  \bar{U}_2 \right) U_1\dif{z}\\
    & +\int_{-\frac{1}{2}}^\frac{1}{2}\left(\Dif\bar{P}_2
      +\left(\Dif^2-k^2\right) \bar{W}_1+\Ray_c \bar{\Theta}_2\right) W_1\dif{z}\\
    &  + \Ray_c \int_{-\frac{1}{2}}^\frac{1}{2} \left(\bar{W}_2
      +\left(\Dif^2-k^2\right)\bar{\Theta}_2 \right)
    \Theta_1\dif{z}\\
    & +\left[\bar{P}_2W_1\right]_{-\frac{1}{2}}^{\frac{1}{2}}
    +\left[\bar{U}_2\Dif{U}_1\right]_{-\frac{1}{2}}^{\frac{1}{2}}
    -\left[U_1\Dif{\bar{U}_2}\right]_{-\frac{1}{2}}^{\frac{1}{2}}
    -\left[\bar{W}_2P_1\right]_{-\frac{1}{2}}^{\frac{1}{2}}\\
    &+\left[\bar{W}_2\Dif{W}_1\right]_{-\frac{1}{2}}^{\frac{1}{2}}
    -\left[W_1\Dif{\bar{W}_2}\right]_{-\frac{1}{2}}^{\frac{1}{2}}
    +\Ray\left( \left[\bar{\Theta}_2\Dif{\Theta}_1\right]_{-\frac{1}{2}}^{\frac{1}{2}}
    -\left[\Theta_1\Dif{\bar{\Theta}_2}\right]_{-\frac{1}{2}}^{\frac{1}{2}}\right)
  \end{split}
\end{equation}
The integral part shows that the adjoint linear system is the same as
the direct one, with $\mbf{L}_c$ as operator. The boundary conditions are the one that allow to
suppress all the boundary values in equation~(\ref{eq:dot-prod-z}). 
The boundary conditions~(\ref{eq:BC-u}) to~(\ref{eq:BC-theta}) are applied to $\mbf{X}_1$ to
remove $\Theta_1(\pm\frac{1}{2})$ and replace $\Dif{U}_1$ and
$P_1$. In addition, the mass conservation equation applied to
$\mbf{X}_2$ allows to replace $\Dif{W}_2$. Factorizing $W_1$, $U_1$
and $\Theta_1$ gives for the boundary conditions
\begin{equation}
  \label{eq:BC-adjoint-tot}
\left[W_1\left(-\bar{P}_2\pm \Phi^\pm
    \bar{W}_2 + 2\Dif{\bar{W}_2} \right)\right]_{-\frac{1}{2}}^{\frac{1}{2}}
+\left[U_1\left(-\ii k \bar{W}_2+\Dif{\bar{U}_2}\right)\right]_{-\frac{1}{2}}^{\frac{1}{2}}
  -\Ray \left[\bar{\Theta}_2\Dif{\Theta}_1\right]_{-\frac{1}{2}}^{\frac{1}{2}}=0.
\end{equation}
Since $W_1$, $U_1$ and $\Dif{\Theta}_1$ can take arbitrary values on
the boundaries, the differences can only be eliminated in a general
manner by setting all their coefficients to 0, which gives the
boundary conditions for the adjoint:
\begin{align}
  \label{eq:BC-u-adj}
  \Dif U_2+\mathrm{i} k W_2 & =0,\\
  \label{eq:BC-w-adj}
  \pm\Phi^\pm W_2 + 2\Dif W_2 - P_2 &=0, \\
  \label{eq:BC-theta-adj}
  \Theta_2 &= 0.
\end{align}
The adjoint problem is therefore identical to the direct one. Among
other implications, all eigenvalues of $L_c$ must be real, which is
consistent with our numerical findings.

\section{Expression of the non-linear terms}
\label{sec:non-lin-term}
Computation of the non-linear term $\mbf{N}(\mbf{X}_n,\mbf{X}_m)$ (eq.~\ref{eq:nonLinear}) is
the trickiest part of the procedure explained in
section~\ref{sec:weakly-non-linear} and deserves some details provided
here. First of all, it contains only a $\mbf{\Theta}$ component, referred to
as $\mbf{N}(\mbf{X}_n,\mbf{X}_m)_\Theta$. To compute it, one needs
first to decompose indices $n$ and $m$ as
\begin{align}
  \label{eq:n-pq}
  n &= 2 p + q \mbox{ with } p = \left\lfloor\frac{n}{2}\right\rfloor,\\
  \label{eq:m-rs}
  m &= 2 r + s  \mbox{ with } r = \left\lfloor\frac{m}{2}\right\rfloor,
\end{align}
where $\lfloor\rfloor$ denotes the floor function. In computing
$\mbf{N}(\mbf{X}_n,\mbf{X}_m)_\Theta$, one needs to account for the
full (i.e. real) expression of $\mbf{X}_n$ and $\mbf{X}_m$ including
the complex conjugate. They write
\begin{align}
  \label{eq:Xn}
  \mbf{X}_n & = \sum_{l_1=0}^{p} \mbf{Z}_{n,2 l_1+q}(z) \ee^{\ii (2 l_1+q) k x}
              + c.c.,\\
  \label{eq:Xm}
  \mbf{X}_m & = \sum_{l_2=0}^{r} \mbf{Z}_{m,2 l_1+s}(z) \ee^{\ii
              (2 l_2+s) k x}
              + c.c..
\end{align}

Using eq.~\eqref{eq:nonLinear}, we get
\begin{equation}
  \label{eq:NXnXmT}
  \begin{split}
    \mbf{N}(\mbf{X}_n,\mbf{X}_m)_\Theta = 
    \sum_{l_1=0}^{p}\sum_{l_2=0}^{q} & \left\{ \left[ \ii (2 l_2 + s) k
        U_{n, 2l_1+q}\Theta_{m, 2 l_2+s}
        \right. \right. \\
        & \left.
        +W_{n, 2 l_1+q} D \Theta_{m, 2 l_2+s} \right]
      \ee^{\ii[2(l_1+l_2)+q+s] k x}\\
    % \right.\\
    & +\left[ - \ii (2 l_2 + s) k U_{n, 2l_1+q}\bar{\Theta}_{m,
          2 l_2+s} \right. \\
        & \left. \left. + W_{n, 2 l_1+q} D \bar{\Theta}_{m, 2 l_2+s} \right]
      \ee^{\ii(2(l_1-l_2)+q-s) k x } \right\} + c.c..
  \end{split}
\end{equation}
The harmonics of the first term is always positive while that of the
second can be negative. Either way, each term has its complex
conjugate and we solve only for the positive or null harmonics, the
rest of the solution simply being obtained as the conjugate of the
computed part.  

% \bibliographystyle{jfm}
% \bibliography{jrn,biblio,bouquin}
% \bibliography{RBPhase}

\end{document}